\documentclass[aps]{revtex4}

\usepackage{amsmath}
\usepackage{natbib}
\usepackage{graphicx}
\usepackage{epstopdf}
\usepackage{subfigure}
\usepackage{dcolumn}
\usepackage{bm}
\usepackage{color}

\begin{document}


\title { Study of chaos in rotating galaxies using extended force-gradient symplectic
methods } 
\author{Li-Na Zhang$^{1,2,3}$} 
\author{Wen-Fang Liu$^{1}$} 
\author{Xin Wu$^{1,2}$}
\email{wuxin_1134@sina.com} 
 \affiliation{1. School of Mathematics,
Physics and Statistics $\&$ Center of Application and Research of
Computational Physics, Shanghai University of Engineering Science,
Shanghai 201620, China \\
2. School of Physical Science and Technology, Guangxi
University, Nanning 530004, China \\
3. School of Physics and Electronics, Hunan Normal University,
Changsha 410081, China}


\begin{abstract}

We take into account the dynamics of three types of models of
rotating galaxies in polar coordinates in a rotating frame. Due to
non-axisymmetric potential perturbations, the angular momentum
varies with time, and the kinetic energy depends on the momenta
and spatial coordinate. The existing explicit force-gradient
symplectic integrators are not applicable to such Hamiltonian
problems, but the recently extended force-gradient symplectic
methods proposed in a previous work are. Numerical comparisons
show that the extended force-gradient fourth-order symplectic
method with symmetry is superior to the standard fourth-order
symplectic method but inferior to the optimized extended
force-gradient fourth-order symplectic method in accuracy. The
optimized extended algorithm with symmetry is used to explore the
dynamical features of regular and chaotic orbits in these rotating
galaxy models. The gravity effects and the degree of chaos
increase with an increase of the number of the
radial terms in the series expansions of the potential. There are
similar dynamical structures of regular and chaotical orbits in
the three types of models for the same number of
the radial terms in the series expansions, energy and initial
conditions.

\textbf{Keywords}: symplectic integrators; chaos; galaxies;
gravity

\end{abstract}



\maketitle%

\section{Introduction}

Chaos in a dynamical system means that the final state of the
dynamical system displays exponentially sensitive dependence on
the initial state. Based on the importance of chaos, many studies
have focused on the subject of chaos in the solar system [1,2] and
galactic dynamics [3-7].

Regular or chaotic motions of particles in galactic dynamics may
affect the fraction of mass. The authors of Refs. [3,5-7]
investigated mass components of non-rotating N-body models of
\emph{elliptical galaxies} in ordered and in chaotic motion.
Voglis et al. [3] found that the fraction of mass in chaotic
motion  is about 24$\%$ of the total mass in one non-rotating
triaxial equilibrium model with smooth centers, and 32$\%$ in
another non-rotating triaxial equilibrium model with smooth
centers. This shows that the spatial distribution of the mass in
chaotic motion is in disagreement with that in ordered motion.
Muzzio et al. [7] pointed  out that the fraction of mass in
chaotic motion of about 53$\%$ in models of non-rotating galaxies
with smooth centers. On the other hand, the fraction of mass in
ordered or chaotic motion was also studied in \emph{spiral
galaxies}. Voglis et al. [8] further showed that rotation leads to
increasing the fraction of mass in chaotic motion (up to the level
of $\approx 65\%$) and shifting the Lyapunov numbers to larger
values in N-body models of rotating galaxies. In other words, the
extent of chaos is substantially enhanced by the rotation, and the
fractions of mass in chaotic motion in the rotating models are
larger than those in the non-rotating models. The spatial
distributions of the dynamical structures along the spiral arms at
the ends of the bar in the barred-spiral galaxy correspond to
those of particles with masses in regular and in chaotic orbits.
In fact, the mass in chaotic motion can almost completely form
spiral arms emanating from the neighborhood of the Lagrangian
points $L_1$ and $L_2$ at the ends of the bar in a barred-spiral
galaxy. The distribution of mass must be associated with
gravitational forces of particles. The gravity is the main driving
mechanism for the formation and the stability of spiral arms in
galaxies. The authors of Refs. [9,10] developed the Moser analytic
series representing the invariant manifolds near the unstable
Lagrangian equilibrium points $L_1$, $L_2$ in a rotating barred
galaxy. In this way, these series can represent the spiral arms,
which are density waves and are composed of chaotic orbits.
Besides the analytical theory, other methods such as the specific
finite time Lyapunov characteristic number, the smaller alignment
index, the surface of section and the frequency analysis were used
to classify the orbits in regular and chaotic cases in barred
galaxies [11-15].

The onset of chaos in two-dimensional Hamiltonian systems of
rotating galaxies in the disc plane in polar coordinates [10] is
due to the inclusion of non-axisymmetric potential perturbations.
These perturbations cause the angular momentum with respect to the
angle to  vary with time and therefore the  Hamiltonian systems
are not integrable. This non-integrability leads to chaos under
some circumstances. Reliable numerical results are always required
to detect the chaotical behavior. In some cases, extremely long
integration times are also required. The adopted computational
schemes for the long-term integration of the Hamiltonian systems
become crucial to reach better stability and higher precisions.
The proper choice of the integrators should naturally be
symplectic schemes, which preserve the symplectic nature of
Hamiltonian dynamics [16,17]. Symplectic methods are a class of
geometric integration algorithms [18] and make the local error in
the total energy not grow with time. There are standard symplectic
methods [19,20] that require some evaluations of the force, and
force-gradient symplectic integrators [16,21-24] that require some
evaluations of force gradient in addition to several evaluations
of the force. Because of good long-term behavior, the standard
symplectic methods have been used in the solar system [1, 25] and
black hole spacetimes [26-31]. They are also suitable for the
two-dimensional Hamiltonian systems of rotating galaxies in polar
coordinates. However, the force-gradient symplectic integrators
are not applicable to such Hamiltonian problems because the
kinetic energies of the Hamiltonian systems depend on the momenta
and spatial coordinates. In fact, they are only adapted to the
integrations of Hamiltonian systems, where the kinetic energies
are quadratic functions of momenta and the potential energies are
functions of coordinates. Energy errors of the force-gradient
symplectic integrators are several orders of magnitude smaller
than those of the standard symplectic algorithms of the same
order, as was confirmed in the literature [22-24]. Recently, our
group extended these force-gradient integrators to the explicitly
integrable kinetic energies, which are not only quadratic
functions of momenta but also depend on coordinates [32]. When the
original force-gradient operator is adjusted appropriately, the
adjusted operator lacks the concept of force gradient and belongs
to the momentum operator like the operator corresponding to the
potential. As a result, the existing explicit force-gradient
symplectic integrators [21-24] are still available in the extended
Hamiltonian systems. The extended force-gradient symplectic
integrators do not alter their symmetry and time reversibility
compared with the existing force-gradient algorithms. The authors
of Ref. [32] used a modified H\'{e}non-Heiles system and a spring
pendulum as two toy models to test the numerical performance of
the extended algorithms. They showed that the fourth-order
Forest-Ruth standard symplectic method does not give true
dynamical properties of order and chaos to the modified
H\'{e}non-Heiles system under some circumstances, whereas the
fourth-order extended force-gradient symplectic methods do. The
obtained results are because the Forest-Ruth method performs
poorer accuracy  than the extended force-gradient algorithms. In
fact, the optimized fourth-order extended force-gradient
symplectic methods have  energy errors that are three orders of
magnitude smaller than those of  the Forest-Ruth method.

Note  that the extended force-gradient symplectic methods proposed
in Ref. [32] were shown to have good numerical performance in the
simulations of the two toy models. Now, we wonder whether the
extended algorithms still exhibit excellent performance in the
application of real astronomical and astrophysical problems. To
evaluate the performance of the extended force-gradient symplectic
methods applicable to the three types of models of rotating barred
galaxies in Refs. [8,10] is one of the main aims in the present
paper. On the other hand, we are interested in studying a
distribution of dynamical structures regarding regular and chaotic
orbits along the radial direction for a given number of the radial
terms of the potential in one of the models using the techniques
of Poincar\'{e} surface of section and fast Lyapunov indicators
(FLIs) [33]. In this way, we desire to know how the number of the
radial terms of the potential affects chaos in one of the models.
We also desire to know what differences in the dynamics exist
among the three types of models with the same number of the radial
terms of the potential, energy and initial conditions.

To implement the three aims, we organize this paper as follows. In
Section 2, we introduce three types of models of rotating barred
galaxies. In Section 3, we apply the extended fourth-order
force-gradient symplectic methods to Model A4 and evaluate the
performance of these algorithms. In section 4, we explore the
dynamical structures in these models using the techniques of
Poincar\'{e} surface of section and FLIs. Finally, the main
results are summarized in Section 5.

\section{Models of rotating barred galaxies}

The motion of a test particle of mass $m_p= 1$ in the plane of a
galaxy with a rotating bar is described in polar coordinates
$(r,\phi)$ in the rotating frame by a two-dimensional Hamiltonian
[8,10]
\begin{equation}
H=\frac{p_{r}^{2}}{2}+\frac{p_{\phi}^{2}}{2 r^{2}}-\Omega
p_{\phi}+\Phi(r, \phi).
\end{equation}
The above notations are given here. $\Omega$ is a pattern speed of
the rotating frame. $p_{r}$ represents a canonical momentum vs the
coordinate $r$, and $p_{\phi}$ is a canonical momentum of the
coordinate $\phi$ corresponding to the angular momentum in the
rest frame. $\Phi(r, \phi)$ denotes the gravitational potential of
the galaxy in the rotating frame and satisfies the Poisson
equation
\begin{equation}
\nabla^2\Phi=4\pi G\rho,
\end{equation}
where $G$ is the constant of gravity and $\rho$ stands for the
density of matter.

A complete expression of the solution of the Poisson equation is
long and complicated. Harsoula et al. [10] took a simple solution
of Eq. (2) as the potential for the $m =2$ mode of the galactic
bar
\begin{equation}
\Phi(r, \phi)=\Phi_{0}(r)+\Phi_{1}(r) \cos 2 \phi+\Phi_{2}(r) \sin
2 \phi,
\end{equation}
where $\Phi_{0}(r)$, $\Phi_{1}(r)$ and $\Phi_{2}(r)$ are functions
of $r$. If $\Phi_{1}(r)=\Phi_{2}(r)=0$, then the potential
$\Phi_{0}(r)=\Phi(r, \phi)$ is axisymmetric and the angular
momentum $p_{\phi}$ is  conserved. When $\Phi_{1}(r)\neq0$ or
$\Phi_{2}(r)\neq0$, the potential $\Phi(r, \phi)$ is
non-axisymmetric and the angular momentum $p_{\phi}$ is not
conserved. Namely, the second and third terms of Eq. (3) act as an
$m =2$ mode of the non-axisymmetric potential perturbation. The
three potential functions are given in Ref. [10] by
\begin{eqnarray}
&&\Phi_{0}(r)= -\frac{1}{R}\left(A_{00}+\frac{1}{4}
A_{20}-\frac{3}{2} A_{22}\right),\\
&&\Phi_{1}(r)=-\frac{3}{2 R}\left(\frac{1}{2} A_{20}+A_{22}\right),\\
&&\Phi_{2}(r)=\frac{3}{2 R} A_{21},
\end{eqnarray}
where $R$ is a numerical constant for the
description of the size of the system. The size is regarded as the
N-body code boundary corresponding to the solutions of the Poisson
equation matched with the solutions of the Laplace equation. The
other notations such as $A_{00}$ are expanded in terms of the
series of spherical Bessel functions $j_{0}$ and $j_{2}$ [34]:
\begin{eqnarray}
A_{00}=\sum_{i=0}^{n}\left[B_{i 00} \cdot j_{0}\left(\xi_{i
0}\right)\right],\\ A_{20}=\sum_{i=0}^{n}\left[B_{i 20} \cdot
j_{2}\left(\xi_{i 2}\right)\right],\\
A_{21}=\sum_{i=0}^{n}\left[C_{i 21} \cdot j_{2}\left(\xi_{i 2}\right)\right],\\
A_{22}=\sum_{i=0}^{n}\left[B_{i 22} \cdot j_{2}\left(\xi_{i
2}\right)\right].
\end{eqnarray}
In the above series expansions of the potential,
$n$ represents the number of the radial terms; that is, each of
the functions like $A_{00}$ in the potential has $(n+1)$ terms.
In addition, $\xi_{il}=a_{il}r/R$. For $l=0$, $a_{i 0}=(i+1/2)\pi$
is the $(i+1)$th root of the equation $j_{-1}(a_{i0})=0$. For
$l=2$, $a_{i2}$ is the solution of the equation $\tan
\left(a_{i2}\right)=a_{i2}$. The solution $a_{i 2}$ should be
restricted in the range $(i-1/2)\pi<a_{i2} <(i+1/2)\pi$ and is
solved by the  Newton iteration method.

The coefficients $B_{i00}$, $B_{i20}$, $C_{i21}$ and $B_{i22}$ in
Eqs. (7)-(10) are obtained through the N-body code on the
positions of the N-body particles and Eq. (2). They are different
for three types of galactic models, which are called as models A,
B and C  in Ref. [10]. The properties of the three models were
described in the papers [8-10]. Here are some of them. Firstly,
the total angular momentum is larger in model B than in model A,
but smaller than in model C. This leads to decreasing the size of
a rotating central bar formed by density waves and increasing the
density in the region of the bar  along the sequence of the models
A, B and C. Secondly, the pattern speed of the bar in each of the
models becomes smaller, whereas the corotation radius gets larger
at the end of a Hubble time. Thirdly, the degree of chaos is
enhanced by rotation increasing the fraction of mass in chaotic
motion. Fourthly, mass in chaotic motion  almost completely
dominates  the formation of spiral arms. Fifthly, invariant
manifolds of all unstable periodic orbits near and beyond
corotation support both the outer edge of the bar and the spiral
arms.  Harsoula provided the coefficients in the three models to
the first author of the present paper via private communication.
The coefficients with $n=19$ are listed in Tables 1-3. The 20
coefficients $B_{i00}$, the 20 coefficients $B_{i20}$ and the 40
coefficients $C_{i21}$, $B_{i22}$ ($i=0,\cdots,19$) in each model
are associated to monopole terms, quadrupole terms and triaxial
terms, respectively. In practice, each group of the coefficients
determine the potential $\Phi(r, \phi)$. Different potentials
$\Phi(r, \phi)$ correspond to different galactic models.
Hereafter, the three models for $n=19$ respectively correspond to
models A19, B19 and C19. Similarly, models  A9, B9 and C9 are
called when $n=9$; models A4, B4 and C4 are also called when
$n=4$.

\begin{table*}[htbp]
\centering \caption{Coefficients of the series (7)-(10) in Model
A.} \label{Tab1}
\begin{tabular}{cccccccccccccc}
\hline
i & 0 &1  & 2 & 3 & 4 & 5 & 6 & 7 & 8 & 9  \\
$Bi00$  & 250670.00 & 214640.00 & 175380.00  & 146580.00 & 126550.00 & 112500.00 & 102250.00 & 93269.00 & 84659.00 & 76738.00 \\
$Bi20$ & 16599.00 & 25348.00 & 25460.00 & 25455.00 & 27770.00 & 31416.00 & 35390.00 & 38187.00 & 39353.00 & 39760.00 \\
$Ci21$ & 616.35 & 1477.80 & 3093.20 & 4784.80 & 5723.80
 & 5186.10 & 3564.00 & 1766.40 & 220.39 & -632.86\\
$Bi22$ & -5808.00 &-9799.20 & -11023.00 & -10369.00 & -8597.30 & -6413.80 & -4878.40 & -4209.20 & -3989.60 & -3932.60\\
\hline
i& 10& 11 & 12 & 13 & 14 & 15 & 16 & 17 & 18 & 19 \\
$Bi00$ & 69503.00 & 62789.00 & 56553.00 & 50735.00 & 45419.00 & 40595.00 & 36128.00 & 31941.00 & 27988.00 & 24271.00\\
$Bi20$ &40303.00 & 41137.00 & 41938.00 & 42164.00 & 41642.00 & 40746.00 & 39745.00 & 38828.00 & 37960.00 & 36816.00\\
$Ci21$ & -833.16 & -816.34 & -786.25 & -741.98 & -684.44 & -601.88
 & -463.25 & -324.97 & -283.78 & -344.23\\
$Bi22$ & -3882.30 & -3768.40 & -3624.80 & -3470.20 &-3355.40 & -3299.00 & -3289.50 & -3298.10 & -3315.100 & -3304.00\\
\hline
\end{tabular}
\end{table*}

\begin{table*}[htbp]
\centering \caption{Coefficients of the series (7)-(10) in Model
B.} \label{Tab2}
\begin{tabular}{cccccccccccccc}
\hline
i & 0 &1  & 2 & 3 & 4 & 5 & 6 & 7 & 8 & 9  \\
$Bi00$ & 249850.00 & 208750.00 & 166240.00 & 136810.00 & 119540.00 & 109650.00 & 102040.00 & 94581.00 & 87464.00 & 80875.00\\
$Bi20$ & 19761.00 & 30778.00 & 32613.00 & 28562.00 & 24516.00 & 25984.00 & 30721.00 & 34872.00 & 37154.00 & 38234.00\\
$Ci21$ & -674.22 & -2590.80 & -4227.30 & -4246.90 & -1974.70 & 118.46 & 87.73 & -1623.40 & -3752.10 &-5087.90\\
$Bi22$ & -6781.10 & -10962.00 & -11493.00 & -9732.80 & -7244.90 & -5062.20 & -3654.80 & -2783.10 & -2305.50 & -2196.80\\
\hline
i& 10& 11 & 12 & 13 & 14 & 15 & 16 & 17 & 18 & 19 \\
$Bi00$ & 74509.00 & 68185.00 & 61997.00 & 56146.00 & 50708.00 & 45630.00 & 40856.00 & 36367.00 & 32180.00 & 28326.00\\
$Bi20$ & 39040.00 & 40062.00 & 41339.00 & 42345.00 & 42784.00 & 42869.00 & 42847.00 & 42584.00 & 41768.00 & 40414.00 \\
$Ci21$ & -5447.70 & -5355.50 & -5293.10 & -5374.40 & -5514.50 & -5533.80 & -5337.50 & -5081.30 & -4882.80 & -4719.30\\
$Bi22$ & -2336.40 & -2490.40 & -2534.30 & -2495.10 & -2474.00 & -2468.80 & -2445.60 & -2417.60 & -2425.40 & -2463.00\\
\hline
\end{tabular}
\end{table*}

\begin{table*}[htbp]
\centering \caption{Coefficients of the series (7)-(10) in Model
C.} \label{Tab3}
\begin{tabular}{cccccccccccccc}
\hline
i & 0 &1  & 2 & 3 & 4 & 5 & 6 & 7 & 8 & 9  \\
$Bi00$  & 249390.00 & 205340.00 & 159420.00 & 129900.00 & 114510.00 & 106340.00 & 100510.00 & 95098.00 & 89366.00 & 83133.00 \\
$Bi20$ & 13998.00 & 21217.00 & 22367.00 & 25113.00 & 29066.00 & 30605.00 & 30582.00 & 31466.00 & 33769.00 & 36355.00\\
$Ci21$ & 89.08 & 1336.80 & 4085.00 & 5662.00 & 3913.40 & 969.22 & -1143.70 & -2174.50 & -2900.50 &-3381.60\\
$Bi22$ & -8969.80 & -14504.00 & -14213.00 & -10170.00 & -5852.70 & -3207.10& -2049.70 & -1654.10 & -1631.90 &-1714.40\\
\hline
i& 10& 11 & 12 & 13 & 14 & 15 & 16 & 17 & 18 & 19 \\
$Bi00$ & 76784.00 & 70724.00 & 65078.00 & 59703.00 & 54546.00 & 49594.00 & 44931.00 & 40471.00 & 36143.00 & 31955.00\\
$Bi20$ & 38644.00 & 40349.00 & 41348.00 & 42023.00 & 42823.00 & 43574.00 & 43919.00 & 43850.00 & 43519.00 & 42874.00\\
$Ci21$ & -3685.00 & -3800.20 & -3575.40 & -3274.30 & -3104.70 & -2945.90 & -2835.40 & -2807.90 & -2761.70 & -2731.70\\
$Bi22$ & -1706.80 & -1606.50 & -1571.10 & -1578.70 & -1567.60 & -1505.90 & -1508.40 & -1617.50 & -1786.00 &-1900.20\\
\hline
\end{tabular}
\end{table*}

The unit systems are those of Refs. [8,10]. The half mass radius
$R_h$ is used as a scaling unit of length, i.e. $r_{scal}= r/R_h$,
where $r$ stands for the real distance and $r_{scal}$ is the
scaling radial distance. For convenience, the scaled radial
distance is still written as $r$ in the later discussions. The
half mass radius is $R_{h}=0.1006$ for Model A, $R_{h}=0.0926$ for
Model B and $R_{h}=0.1167$ for Model C. The size of the system is
$R=0.85$. Here, one unit of length is 8kpc. The unit of time is
the half mass crossing time $T_{hmct}=[2R^3_h/(GM)]^{1/2}
=t_{Hub}/300$, where $t_{Hub}$ represents a Hubble time. The
pattern speeds of the rotating frame in the three models A, B and
C are taken as $\Omega_{A} = 5886.65$, $\Omega_{B}=6010.36$ and
$\Omega_{C} = 6137.14$, which correspond to $20 \sim 25$ km
sec$^{-1}$ kpc$^{-1}$ in real units.

As is mentioned above, the second and third terms of Eq. (3)
destroy the axial-symmetry of the system (1) such that the angular
momentum $p_{\phi}$ varies with time. Thus, no additional
constants of motion but the Hamiltonian $H$ of Eq. (1) exists.
This implies the nonintegrability of the system (1) with two
degrees of freedom in a four-dimensional phase space. A numerical
method is a good tool to solve this nonintegrable problem.

\section{A choice of numerical integrator}

A prior choice to an integrator for a long-term integration of the
Hamiltonian (1) is naturally a symplectic method with the
conservation of the Hamiltonian flow. Following this idea, we
consider the application of symplectic integration to this
Hamiltonian problem.

\subsection{Generalized force-gradient symplectic integrators}

For convenience, we take $\mathbf{p}=(p_{1}, p_{2})=(p_{r},
p_{\phi})$ and $\mathbf{q}=(q_{1}, q_{2})=(r, \phi)$. We rewrite
the Hamiltonian (1) as
\begin{equation}
H(\mathbf{p}, \mathbf{q})=T(\mathbf{p}, q_1)+\Phi(\mathbf{q}),
\end{equation}
where the kinetic energy $T$ is a function of the momenta
$\mathbf{p}$ and coordinate $r$
\begin{equation}
T(\mathbf{p},q_1)=\frac{p_{r}^{2}}{2}+\frac{p_{\phi}^{2}}{2
r^{2}}-\Omega p_{\phi}.
\end{equation}
The lie derivative operators of $T$ and $\Phi$ are defined as
\begin{eqnarray}
\mathcal{A} &=& \{, T\}=\sum_{i=1}^{2}\left(T_{p_{i}}
\frac{\partial}{\partial q_{i}}-T_{q_{i}} \frac{\partial}{\partial
p_{i}}\right) = p_{r} \frac{\partial}{\partial r}+\left(\frac{
p_{\phi}}{r^2}-\Omega \right)\frac{\partial}{\partial \phi}
+\frac{p^2_{\phi}}{r^3} \frac{\partial}{\partial p_{r}}, \\
\mathcal{B} &=& \{, \Phi\}=-\frac{\partial \Phi}{\partial r}
\frac{\partial}{\partial p_{r}}-\frac{\partial \Phi}{\partial
\phi} \frac{\partial}{\partial p_{\phi}},
\end{eqnarray}
where the symbols $\{,\}$ represent the Poisson brackets.
Obviously, $T$ and $\Phi$ can be exactly, analytically solvable,
and $\mathcal{A}$, $\mathcal{B}$ correspond to their solvers.

The solvers $\mathcal{A}$ and $\mathcal{B}$ can symmetrically
compose a second-order symplectic leapfrog integrator
\begin{eqnarray}
\mathrm{M}2 &=& e^{\frac{\tau}{2}\mathcal{A}} e^{\tau
\mathcal{B}}e^{\frac{\tau}{2}\mathcal{A}} =
e^{\tau(\mathcal{A}+\mathcal{B})-\frac{\tau^3}{12}
[\mathcal{B},[\mathcal{A}, \mathcal{B}]]+\frac{\tau^3}{24}
[\mathcal{A},[\mathcal{B}, \mathcal{A}]]},
\end{eqnarray}
where $\tau$ is a time step, and two commutators are
\begin{eqnarray}
\mathcal{C} &=& [\mathcal{B},[\mathcal{A},
\mathcal{B}]]=[\mathcal{B}, \mathcal{A} \mathcal{B}-\mathcal{B}
\mathcal{A}] = 2\mathcal{B}\mathcal{A}\mathcal{B}
-\mathcal{B}\mathcal{B}\mathcal{A}-\mathcal{A}
\mathcal{B}\mathcal{B}, \\
\mathcal{D} &=& [\mathcal{A},[\mathcal{B}, \mathcal{A}]]=2
\mathcal{A}\mathcal{B}\mathcal{A}-\mathcal{A}\mathcal{A}\mathcal{B}-\mathcal{B}
\mathcal{A}\mathcal{A}.
\end{eqnarray}
In the second line of Eq. (15), the first term
$\tau(\mathcal{A}+\mathcal{B})$ corresponds to the numerical
solution of the Hamiltonian (1), and the second and third terms
correspond to local truncation errors of the numerical solution
remaining at an order $\mathcal{O}(\tau^3)$. Because of such local
truncation errors, the algorithm $\mathrm{M}2$ can give a
second-order accuracy to the explicit numerical solution. In terms
of the solvers $\mathcal{A}$ and $\mathcal{B}$, a fourth-order
symplectic method of Forest and Ruth [19] is established by
\begin{eqnarray}
\mathrm{M} 4 &=& e^{\alpha \tau \mathcal{A}} e^{\beta \tau
\mathcal{B}} e^{\left(\frac{1}{2} -\alpha\right) \tau \mathcal{A}}
e^{(1-2 \beta) \tau \mathcal{B}}
e^{\left(\frac{1}{2}-\alpha\right) \tau \mathcal{A}} e^{\beta \tau
\mathcal{B}} e^{\alpha \tau \mathcal{A}},
\end{eqnarray}
where $\beta=1 /(2-\sqrt[3]{2})$ and $\alpha=\beta / 2$.

If $r^2$ in the second term  of Eq. (12) is absent, $\mathcal{A}$
of Eq. (13) is a position operator which acts on only the position
coordinates. $\mathcal{C}=2\mathcal{B}\mathcal{A}\mathcal{B}$ is
similar to the momentum operator $\mathcal{B}$, which acts on only
the momenta. In fact, $\mathcal{C}$ is an exactly, analytically
solvable operator corresponding to the force gradient of the
gravitational potential. However, $\mathcal{D}$ is a momentum and
position mixed operator and is not  an exactly analytical solver.
In view of these facts,  a composition of $\mathcal{B}$ and
$\mathcal{C}$ appears in a class of explicit symplectic
integrators, called the force-gradient symplectic integration
algorithms [16,21-24].

Now, $r^2$ in the second term of Eq. (12) remains. Although
$\mathcal{A}$ of Eq. (13) is a momentum and position mixed
operator, it is still  an exactly analytical solver in this case.
Is $\mathcal{C}$ is an exactly analytical solver? We said yes as
an answer to this question in our previous work [32]. In fact, the
answer can be given through the operators acting on the momenta
and positions in the second line of Eq. (16). It is easy to check
that $\mathcal{A} \mathcal{B} \mathcal{B} q_{i}=\mathcal{A}
\mathcal{B} \mathcal{B} p_{i} \equiv 0$, $\mathcal{B} \mathcal{A}
\mathcal{B} q_{i} \equiv 0$, $\mathcal{B} \mathcal{B} \mathcal{A}
q_{i} \equiv 0$, and
\begin{equation}
\mathcal{B} \mathcal{A} \mathcal{B} p_{i}=\sum_{j=1}^{2}
\sum_{k=1}^{2} \Phi_{q_{i} q_{j}}\Phi_{q_{k}} T_{p_{j} p_{k}},
\end{equation}
\begin{equation}
\mathcal{B} \mathcal{B} \mathcal{A} p_{i}=-\sum_{j=1}^{2}
\sum_{k=1}^{2} \Phi_{q_{j}} \Phi_{q_{k}} T_{q_{i} p_{j} p_{k}},
\end{equation}
where $\Phi_{q_{j}}=\partial \Phi / \partial q_{j}$,
$\Phi_{q_{k}}=\partial \Phi / \partial q_{k}$, and $\Phi_{q_{i}
q_{j}}=\frac{\partial^{2} \Phi}{\partial q_{i} \partial q_{j}}$,
$T_{p_{j} p_{k}}=\frac{\partial^{2} T}{\partial p_{j} \partial
p_{k}}$, $T_{q_{i} p_{j} p_{k}}=\frac{\partial^{3} T}{\partial
q_{i}
\partial p_{j} \partial p_{k}}$. Clearly, $\mathcal{C}$ is also a
momentum operator:
\begin{eqnarray}
  \mathcal{C} = \nonumber 2\mathcal{B}\mathcal{A}\mathcal{B}-\mathcal{B}
  \mathcal{B}\mathcal{A} = \sum_{i=1}^{2} \sum_{j=1}^{2} \sum_{k=1}^{2}(2 \Phi_{q_i q_j} \Phi_{q_k}
  T_{p_jp_k}  +\Phi_{q_j} \Phi_{q_k} T_{q_i p_j p_k}) \frac{\partial}{\partial
  p_{i}}.
\end{eqnarray}
The momentum operator $\mathcal{C}$ of Eq. (21) is no longer the
force gradient of the gravity potential, but can be regarded as an
extension of the force-gradient factor in Refs. [16,21-24].
Without doubt, it is an exactly analytical solver similar to the
momentum operator $\mathcal{B}$.

Symmetric composition methods of the three operators
$\mathcal{A}$, $\mathcal{B}$ and $\mathcal{C}$ can yield various
explicit symplectic algorithms. For instance, a five-stage
fourth-order symplectic method including the operator
$\mathcal{C}$ was proposed by Chin [22] in the form
\begin{eqnarray}
\mathrm{N} 4 =\nonumber
e^{\frac{\tau}{2}\left(1-\frac{1}{\sqrt{3}}\right) \mathcal{A}}
e^{\frac{\tau}{2}\left(\mathcal{B}+\frac{\tau^{2}}{24}(2-\sqrt{3})
\mathcal{C}\right)} e^{\frac{\tau}{\sqrt{3}} \mathcal{A}}
e^{\frac{\tau}{2}\left(\mathcal{B}+\frac{\tau^{2}}{24}(2-\sqrt{3})
\mathcal{C}\right)}
e^{\frac{\tau}{2}\left(1-\frac{1}{\sqrt{3}}\right) \mathcal{A}}.
\end{eqnarray}
Omelyan et al. [24] also constructed a seven-stage fourth-order
optimized symplectic algorithms with the inclusion of the operator
$\mathcal{C}$:
\begin{eqnarray}
\mathrm{N} 4 \mathrm{P} = e^{\theta \tau \mathcal{A}} e^{\lambda
\tau\left[\mathcal{B}+(2 \xi+\chi) \tau^{2} \mathcal{C}\right]}
e^{(1-2 \theta) \frac{\tau}{2} \mathcal{A}} e^{(1-2 \lambda)
\tau\left[\mathcal{B}+(2 \xi+\chi) \tau^{2} \mathcal{C}\right]}
e^{(1-2 \theta) \frac{\tau}{2} \mathcal{A}} e^{\lambda
\tau\left[\mathcal{B}+(2 \xi+\chi) \tau^{2} \mathcal{C}\right]}
e^{\theta \tau \mathcal{A}},
\end{eqnarray}
where the time coefficients are
\begin{eqnarray}
&&\theta=0.1159953608486416 \times 10^{0}, \nonumber \\ \nonumber
&&\lambda=0.2825633404177051 \times 10^{0}, \\ \nonumber
&&\chi=0.3035236056708454 \times 10^{-2}, \\ \nonumber
&&\xi=0.1226088989536361 \times 10^{-2} .
\end{eqnarray}
The concept of optimization is the time coefficients minimizing
the norm of the leading term of fifth-order truncation errors.

The construction mechanisms of the  algorithms with the extended
operator $\mathcal{C}$ was discussed in the previous work [32]. In
fact, the extended force-gradient algorithms acting on the
original system (1) are equivalent to the standard symplectic
methods (like the method M4 without the extended operator
$\mathcal{C}$) acting on the modified Hamiltonian systems.

\subsection{Numerical tests}

Model A4  is used  as a test model to evaluate the performance of
the algorithms M4, N4 and N4P. For comparison, an eighth- and
ninth-order Runge-Kutta-Fehlberg integrator [RKF8(9)] with
adaptive step sizes is also employed. The time step is
$\tau=0.0005\times T_{hmct}=0.0005\times
t_{hub}/300=0.0005/82.4/300\approx 2.0\times 10^{-8}$
(sec$\cdot$Mpc/km)=$2.0\times 10^{-5}$ (sec$\cdot$kpc/km)
$\approx$ $2.0\times 10^{-5}$ ($9.5\times 10^{8}$ y) = $0.19\times
10^{5}$ y. In short, $\tau\approx2.0\times 10^{-5}$
(sec$\cdot$kpc/km) is used in our codes. The authors of Ref. [8]
pointed out that the time step is a good choice because the
hysteresis between the bar of the potential and the bar of the
real density of particles during the time is minimal and the
cumulative effect of a numerical retarding torque in a Hubble time
is small.

The energy of the system  (1) is $E =H= -7\times10^{6}$. The
initial conditions are $r= 0.164/R_h=0.164/0.1006$, $\phi =
1.89\pi$, $p_{r} = 0$. The initial momentum $p_{\phi}>0$ is
determined by $E = H$. Fig. 1(a) plots the relative energy errors
$|\Delta H/H_{0}| = |(E_{t} - E)/E|$ for these several algorithms,
where $E_{t}$ is the numerical energy at time $t$. Here, each
algorithm has $2.5\times 10^{8}$ steps, which correspond to time
$t\approx 5056.63$ (sec$\cdot$kpc/km). The three symplectic
methods M4, N4 and N4P show no secular growth in the energy
errors. This is an inherent advantage of these symplectic methods.
Among the three symplectic integrators, the standard fourth-order
symplectic method M4 without the extended operator $\mathcal{C}$
performs the poorest accuracy, whereas the fourth-order optimized
symplectic algorithm N4P with the extended force-gradient operator
$\mathcal{C}$ exhibits the best accuracy. These numerical results
show that the inclusion of the extended force-gradient operator
has an advantage over the exclusion of the extended force-gradient
operator in the accuracy. The  optimized method is also better
than the corresponding non-optimized method. Unlike these
symplectic methods, the non-symplectic integrator RKF8(9) makes
the energy error grow with time.  RKF8(9) is superior to N4P in
the accuracy, but inferior  to N4P in the computational
efficiency, as is shown in Table 4. Taking the solutions of
RKF8(9) as reference solutions, we obtain the relative position
errors of the three symplectic algorithms M4, N4 and N4P in Fig.
1(b). As a result, the method M4  still has the largest position
error, while the method N4P yields the smallest one. When many
other values of the energy and initial conditions are also used,
they  do not affect the numerical performances of these
integrators. In fact, the numerical performance of an integrator
is independent of a choice of the parameters and initial
conditions [16-32]. In spite of this, it is
necessary to choose bounded orbits as testing the integrator's
performance. Thus, an appropriate choice of the parameters and
initial conditions is still necessary.

Because of the fourth-order optimized symplectic algorithm N4P
with the extended force-gradient operator $\mathcal{C}$ having the
best performance in the stabilization of energy errors, it is used
to study the dynamics of orbits.

\begin{table*}[htbp] \centering \caption{Relative energy
errors, relative radial errors and CPU times for the algorithms in
Fig. 1. Note that -12.2 denotes the error with an order of
$10^{-12.2}$, and -2.1 means the error being an order of
$10^{-2.1}$. $17'42''$ in CPU times means 17 minutes and 42
seconds.} \label{Tab4}
\begin{tabular}{ccccc}
\hline
Method  & RKF8(9) & M4 & N4 & N4P \\
\hline $|\Delta H/E|$ & -12.2 & -8.6  & -9.5 & -11.0 \\
\hline $|\Delta r/r_0|$ &  / &  -2.1 & -2.9  & -3.4   \\
\hline Time & $17'42''$  & $4'30''$  & $5'43''$  &  $6'57''$  \\
\hline
\end{tabular}
\end{table*}

\section{Regular and chaotic dynamics}

The authors of  Ref. [10] applied the Moser
theory of invariant manifolds to the chaotic spiral arms. The
invariant manifolds starting at the Lagrangian points $L_1$,
$L_2$, or unstable periodic orbits around $L_1$ and $L_2$ are
described in terms of series. The convergence of the series is an
analytical means for studying chaotic orbits with initial
conditions in the neighborhood of the invariant manifolds of the
unstable point $L_1$ or $L_2$. A domain of the convergence of the
series around every Lagrangian point is a Moser domain. The
intersection of the orbits inside the Moser domain with an
apocentric section produces the spiral structure of this domain.
The chaotic orbits with initial conditions near, but in the
exterior of the boundary of the Moser domain of convergence become
chaotic attractors. Such chaotic orbits are also identified by the
techniques of Poincar\'{e} sections and fast Lyapunov indicators
(FLIs) [33]. The method of Poincar\'{e} sections, which shows
intersections of the particles' trajectories with the surface of
section in phase space, can clearly describe the phase space
structure of a conservative 4-dimensional system. One or several
isolated points on the Poincar\'{e} sections correspond periodic
orbits. Kolmogorov-Arnold-Moser (KAM) tori on the Poincar\'{e}
sections correspond regular quasi-periodic orbits. If there are
many plotted points that are distributed randomly in an area, the
motion is chaotic. In a word, the distribution of the points in
the Poincar\'{e} map can show whether or not the motion is
chaotic. For ordered and chaotic  orbits, the length of a
deviation vector increases in completely different time rates. The
FLI uses the  completely different time rates to distinguish
between the ordered and chaotic cases. The Moser theory of
invariant manifolds for allowing to study chaotic orbits requires
that the series describing the Hamiltonian dynamics near an
unstable equilibrium point or an unstable periodic orbit be
convergent. It may not work well for the non-convergent Birkhoff
normal form series around stable invariant points, or stable
periodic orbits. The techniques of Poincar\'{e} sections and FLIs
for finding chaos do not have this restriction. Clearly, they can
detect chaotic orbits in larger regions compared with the Moser
theory of invariant manifolds. Here, we employ the techniques of
Poincar\'{e} sections and FLIs to investigate the dynamical
behavior of the three types of bar spiral galaxy models A, B and
C. The considered orbits are not restricted to those around the
unstable Lagrangian points $L_1$ and $L_2$. We are mainly
interested in comparing the differences in the dynamical behavior
among these types of models. The effect of the number $(n+1)$ of
the radial terms in the series expansions of the potential on the
dynamical behavior in each set of models is also considered.

\subsection{Models A4, A9 and A19}

The parameters and initial conditions are those of Fig. 1, but the
initial values of $r$ and $p_{\phi}$ are different. Model A4
exhibits different phase space structures in two ranges of the
initial separations $r$, as is shown through Poincar\'{e} sections
at the  plane $\phi=\pi/2$ with $p_{\phi}>0$ in Figs. 2 (a) and
(b). The three orbits in Fig. 2(a) are KAM tori, which correspond
to the regularity of the orbits. One orbit with the initial
separation $r=4.430/0.1006$ and another orbit with the initial
separation $r=4.424/0.1006$ in Fig. 2(b) seem to have no explicit
difference from the phase space structures on the section, but
they exhibit different regular and chaotic dynamical features,
which can be identified clearly by means of the technique of FLI
in Fig. 2(c). The FLI is that with two nearby orbits defined in
Ref. [35]:
\begin{eqnarray}
\textrm{FLI}=\log_{10}\frac{d(t)}{d(0)},
 \end{eqnarray}
where $d(0)$ and $d(t)$ are the distances between two nearby
orbits at time 0 and $t$, respectively. The orbit with the initial
separation $r=4.430/0.1006$ corresponds to the FLI increasing in a
power law with time $\log_{10}t$ and is regular. However, the
orbit with the initial separation $r=4.424/0.1006$ has the FLI
increasing in an exponential law with time and should be chaotic.
After $5\times 10^{4}$ integration steps, the FLIs with completely
different increasing laws with time are sensitive to distinguish
the two cases of order and chaos.

The phase space structures for Model A9 in Fig. 3(a) are somewhat
unlike those for Model A4 in Fig. 2. When the initial separation
$r$ is given in a small range of $1.45<r<1.7$ for Model A4, the
orbits are regular KAM tori (not plotted). However, the orbit with
the initial separation $r=0.153/0.1006$ in Model A9 has several
hyperbolic points. Its chaoticity is confirmed by the FLI of Fig.
3(c). There are no chaotic orbits in a range of $10<r<12.3$ for
Model A4 in Fig. 2(a), but there are two chaotic orbits with the
initial separations $r=1.210/0.1006$ and $r=1.235/0.1006$ in the
range of $10<r<12.3$ for Model A9 in Fig. 2 (b) and  (c).

When Model A19 is considered in Fig. 4(a), an orbit with the
initial separation $r=0.163/0.1006$ is a regular KAM torus.
Another orbit with the initial separation $r=0.154/0.1006$
consists of many islands and therefore is still regular. However,
the third orbit with the initial separation $r=0.160/0.1006$ is
chaotic because it has many discrete points which are randomly
filled with an area. The degree of chaos seems to be strengthened
in the small range of $1.45<r<1.7$ from A4 to A9 and to A19. Chaos
in the range of $10<r<12.5$ seems to be stronger for A19 in Fig.
4(b) than  for A9 in Fig. 3(b). When the initial separation $r$
spans from 20 to 23 in Fig. 2(c), a lot of regular KAM torus
orbits exist  in the interior of the ordered orbit with the
initial separation $r=2.230/0.1006$. There is a small chaotic
region between the two orbits of the initial separations
$r=2.230/0.1006$ and $r=2.290/0.1006$. The regularity of the orbit
with the initial separation $r=2.230/0.1006$ and the chaoticity of
the orbit with the initial separation $r=2.290/0.1006$ can be
identified clearly by means of the technique of FLIs in Fig. 4(d).

Seen from Figs. 2-4, the number and degree of chaotic orbits seem
to increase as the number of the radial terms $n$ gets large. This
result is clearly shown in Fig. 5 that lists the dependence of FLI
on the initial separation $r$ in each of Models A4, A9 and A19. In
other words, Fig. 5 shows that a distribution of the phase space
structures of regular and chaotic orbits for each model depends on
a range of the initial separation $r$. Each of the FLIs is
obtained after $5\times 10^{4}$ integration steps. 7 is a critical
value of the FLI between the ordered and chaotic two cases. The
FLIs not more than 7 indicate the regularity of bounded orbits,
whereas those larger than 7 correspond to the chaoticity of
bounded orbits. The result on an increase of $n$ increasing the
number and degree of chaotic orbits is based on $n$ corresponding
to the number of the radial terms in the series expansions of the
potential. An increase of $n$ means that of the number of the
radial terms. Equivalently, the nonlinear gravitational
interaction effects in the potential are enhanced.

\subsection{Models B4, B9 and B19}

Fig. 6 describes the phase space structures Model B4  in two
ranges of the initial separations $r$. There is a weak chaotic
orbit with the initial separation $r=0.750/0.0926$ in the range of
$7.2<r<8.2$ in Fig 6(a). The two orbits in the range of
$10.0<r<12.2$ are regular KAM tori in Fig 6(b). However, chaos
occurs for the initial separation  $r=3.746/0.0926$ in the range
of $40.25<r<43$ in Fig. 4(c). The dynamical features of the
chaotic orbit in panel (a) and the two orbits in panel (c) are
also shown in Fig. 6(d). The phase space structures of Model B4 in
Fig. 6 (b) and (c) are similar to those of  Model A4 in Fig. 2 (a)
and (b).

The phase space structures of Model B9 in Fig. 7 (a) and (b) also
look like those of Model A9 in Fig. 3 (a) and (b). The initial
separations $r=0.157/0.0926$ and $r=0.139/0.0926$ correspond to
two regular orbits consisting of three loops in Fig. 7(a). Chaos
is present for $r=0.141/0.0926$ in Fig. 7(a) and $r=1.128/0.0926$
in Fig. 7(b), as the FLIs of Fig. 7(c) show.

Model B19  seem to have stronger chaos in the ranges of
$1.425<r<1.7$ and $10<r<12.25$ in Fig. 8 (a) and (b) than Model B9
in Fig. 7 (a) and (b). For the initial separation $r$ belonging to
the range of $1.425<r<1.7$, the orbital dynamical behavior for
Model B19 in Fig. 8(a) is greatly different from that for Model
A19 in Fig. 4(a). However, the phase space structures of Model B19
in Fig. 8 (b) and (c) also resemble those of Model A19 in Fig. 4
(b) and (c). The FLIs of Fig. 8(d) support the chaoticity of the
two orbits for $r=2.368/0.0926$ and $r=2.420/0.0926$ in Fig. 8(c).

The relation between the FLI and the initial separation $r$ in
Fig. 9 clearly shows that the number and degree of chaotic orbits
in Models B increase with $n$ increasing.

\subsection{Models C4, C9 and C19}

As far as Models C are concerned, chaos seems to get stronger in
the range of $3.5<r<5.1$ from $n=4$ in Fig. 10(a) to $n=9$ in Fig.
10(c) and $n=19$ in Fig. 10(e). The result is also suitable for
the range of $10<r<12.2$ in Fig. 10 (b), (d) and (f). The relation
between the FLI and the initial separation $r$ in Fig. (11)
supports that chaos becomes stronger with $n$ increasing.

Several points can be concluded from the above demonstrations. The
three types of models A, B and C have the same expressions, but
have differences in the pattern speeds $\Omega$ and the
coefficients $B_{i00}$, $B_{i20}$, $C_{i21}$ and $B_{i22}$. The
three models have similar dynamical behaviors when they have same
radial term number $n$, energy $E$ and initial conditions. The
dynamical structures of ordered and chaotic orbits in each model
have different distributions along the radial direction. When the
radial term number $n$ increases, the gravity effects are
enhanced. This leads to strengthening the degree of chaos. It is
worth noting that the main structures of the spiral arms in the
models A, B and C are formed at radial distances between 3 and 5
scaled radii, as is shown in Refs. [8] and [10]. Some  ranges of
the scaled radial distances, such as the range of the scaled
radial distances from 40 to 43, seem to be very far from the main
structures of the spiral arms. These results show that chaos can
occur at the radial distances close to the main structures of the
spiral arms and those far from the main structures of the spiral
arms. In addition, the chaotic orbits we find are
not restricted in the regions near the unstable equilibrium points
$L_1$ and $L_2$. Thus, the energies selected in this paper are
unlike those of [8-10].

\section{Conclusions}

This paper mainly focuses on three classes of models of rotating
galaxies in the polar coordinates in the rotating frame. They are
models A, B and C, which have the same expressions but have
different pattern speeds $\Omega$ and different coefficients
$B_{i00}$, $B_{i20}$, $C_{i21}$ and $B_{i22}$. When the potentials
with large deviations from axial symmetry are included in these
models, the angular momentum $p_\phi$ is not a constant of motion.
In this case, the kinetic energy is a function of the momenta and
spatial coordinate.

The existing explicit force-gradient symplectic integrators
[21-24] do not work for the present Hamiltonian problems. However,
our extended force-gradient symplectic methods in the previous
work [32] are still available. Numerical tests show that the
fourth-order symmetric symplectic method without the extended
force-gradient operator performs poorer accuracy than that with
the extended force-gradient operator. In particular, the optimized
extended force-gradient symplectic method is superior to the
corresponding non-optimized one in accuracy.

The fourth-order optimized symplectic algorithm with the extended
force-gradient operator is applied to survey the dynamical
features of regular and chaotic orbits in these rotating galaxy
models. It is shown through the techniques of Poincar\'{e}
sections and fast Lyapunov indicators that an increase of the
radial term number of the potential strengthens the gravity
effects and the degree of chaos. The three types of models have
similar dynamical structures for the same radial term number,
energy and initial conditions.

\textbf{Author Contributions}: Software and Writing-original
draft, L. Z.; Methodology, W. L.; Supervision, Conceptualization,
Writing - Review $\&$ Editing and Funding Acquisition, X. W. All
authors have read and agreed to the published version of the
manuscript.

\textbf{Funding}: This research has been supported by the National
Natural Science Foundation of China (Grant Nos. 11973020 and
11533004), and the Natural Science Foundation of Guangxi (Grant
No. 2019GXNSFDA245019).

\textbf{Data Availability Statement}: Our paper is a theoretical
work. All of the data are calculated and given in the paper.

\textbf{Institutional Review Board Statement}: Not applicable.

\textbf{Informed Consent Statement}: Not applicable.

\textbf{Acknowledgments}: The first author is very grateful to Dr.
Harsoula for the series coefficients in Tables 1-3. The authors
also thank anonymous referees for useful suggestions.

\textbf{Conflicts of Interest}: The authors declare no conflict of
interest.

\newpage

\begin{figure*}
\center{
\includegraphics[scale=0.36]{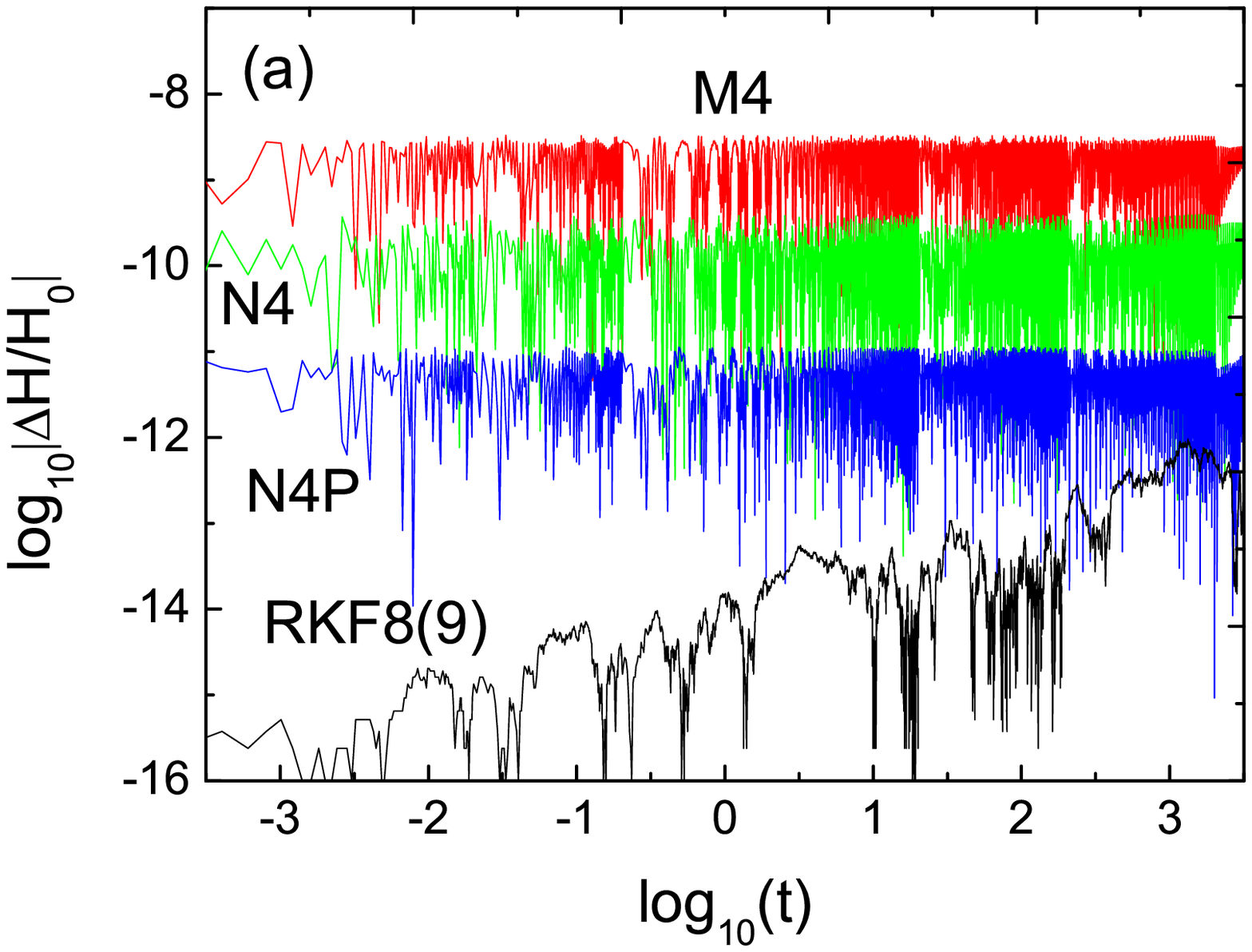}
\includegraphics[scale=0.36]{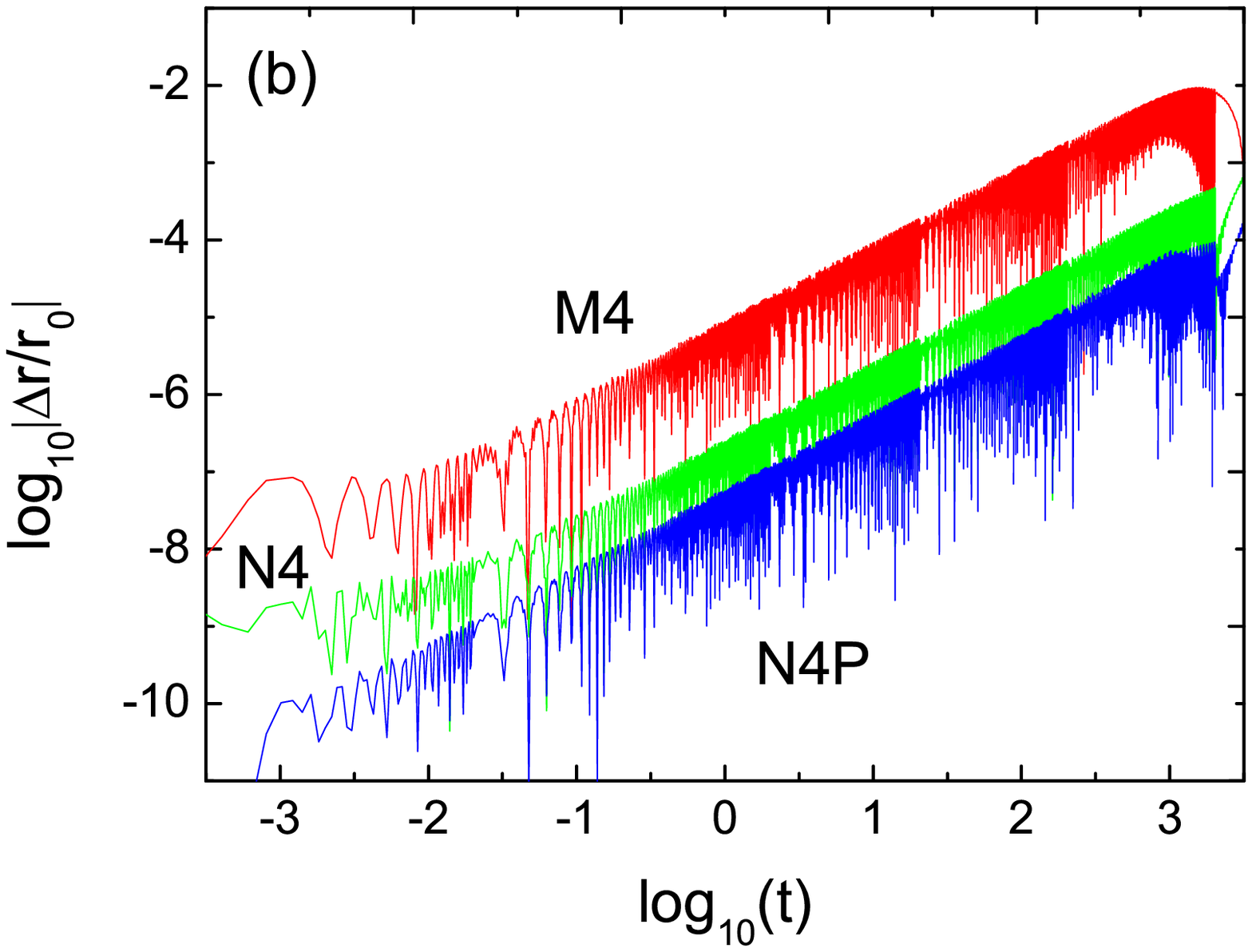}
\caption{(a) The relative energy errors $\Delta H/H_0=(E_t-E)/E$
for the standard fourth-order symplectic method M4, the extended
force-gradient fourth-order symplectic method N4,  the optimized
extended force-gradient fourth-order symplectic method N4P and the
non-symplectic integrator RKF8(9) independently acting on Model
A4. $E_t$ is the numerical energy of the system (1) at time $t$,
and the energy is $E=-7\times10^{6}$. The time step is
$\tau=0.0005\times T_{hmct}\approx 2\times 10^{-5}$
(sec$\cdot$kpc/km). The initial conditions are $r=0.164/R_h=
0.164/0.1006$, $\phi=1.89\pi$ and $p_{r}=0$. The positive initial
value of $p_\phi$ is determined by $E=H$. The three symplectic
methods do not give secular drifts to the energy errors, whereas
the non-symplectic integrator RKF8(9) does. In accuracy, M4 is the
poorest one among these symplectic integrators, while N4P is the
best one. (b) The relative position error $\Delta
r/r_0=(r-r_0)/r_0$ between the radial separation $r_0$ given by
RKF8(9) and the radial separation $r$ given by one of the
symplectic methods. The position error for N4 is smaller than for
M4 but larger than for N4P. }} \label{f1}
\end{figure*}

\begin{figure*}
\center{
\includegraphics[scale=0.19]{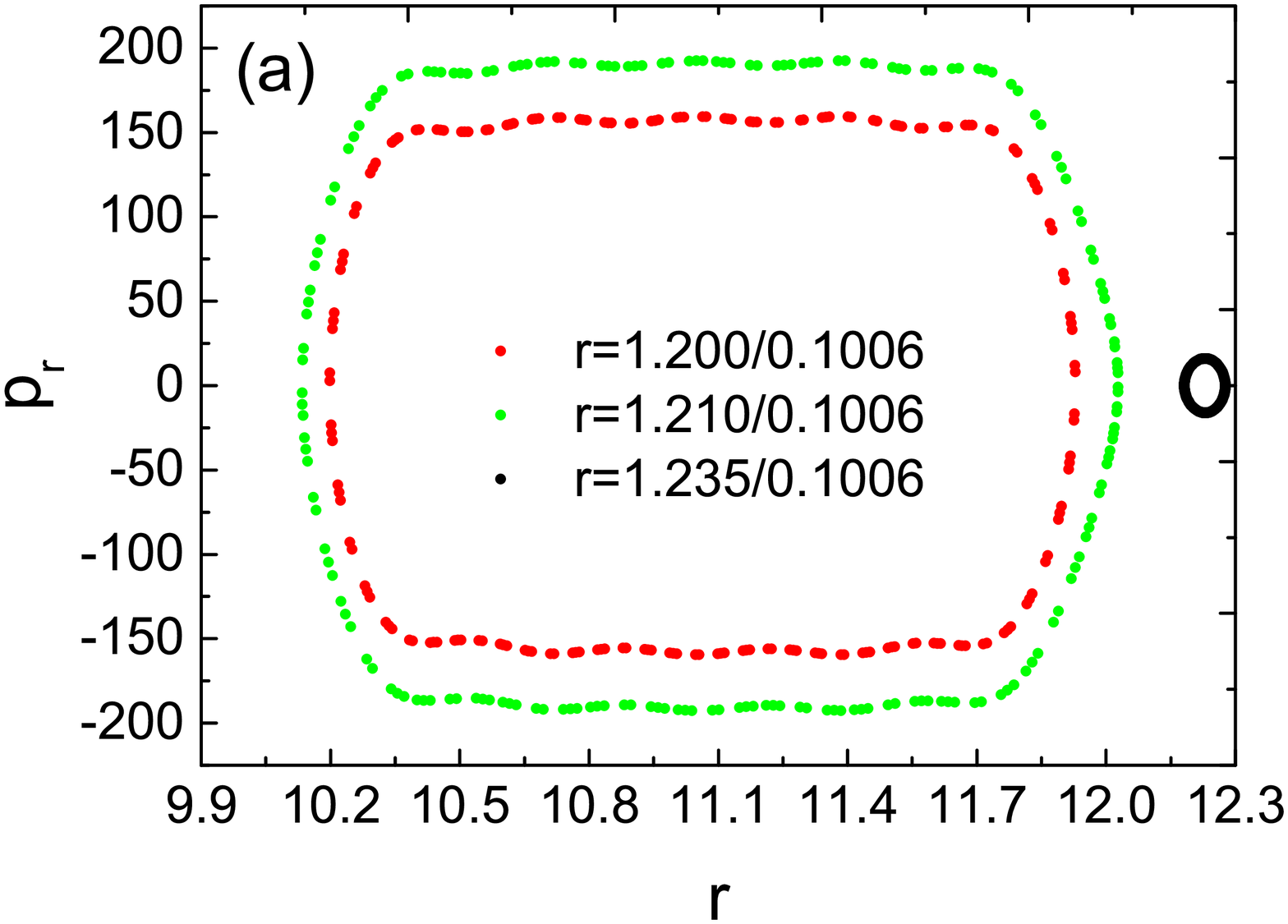}
\includegraphics[scale=0.19]{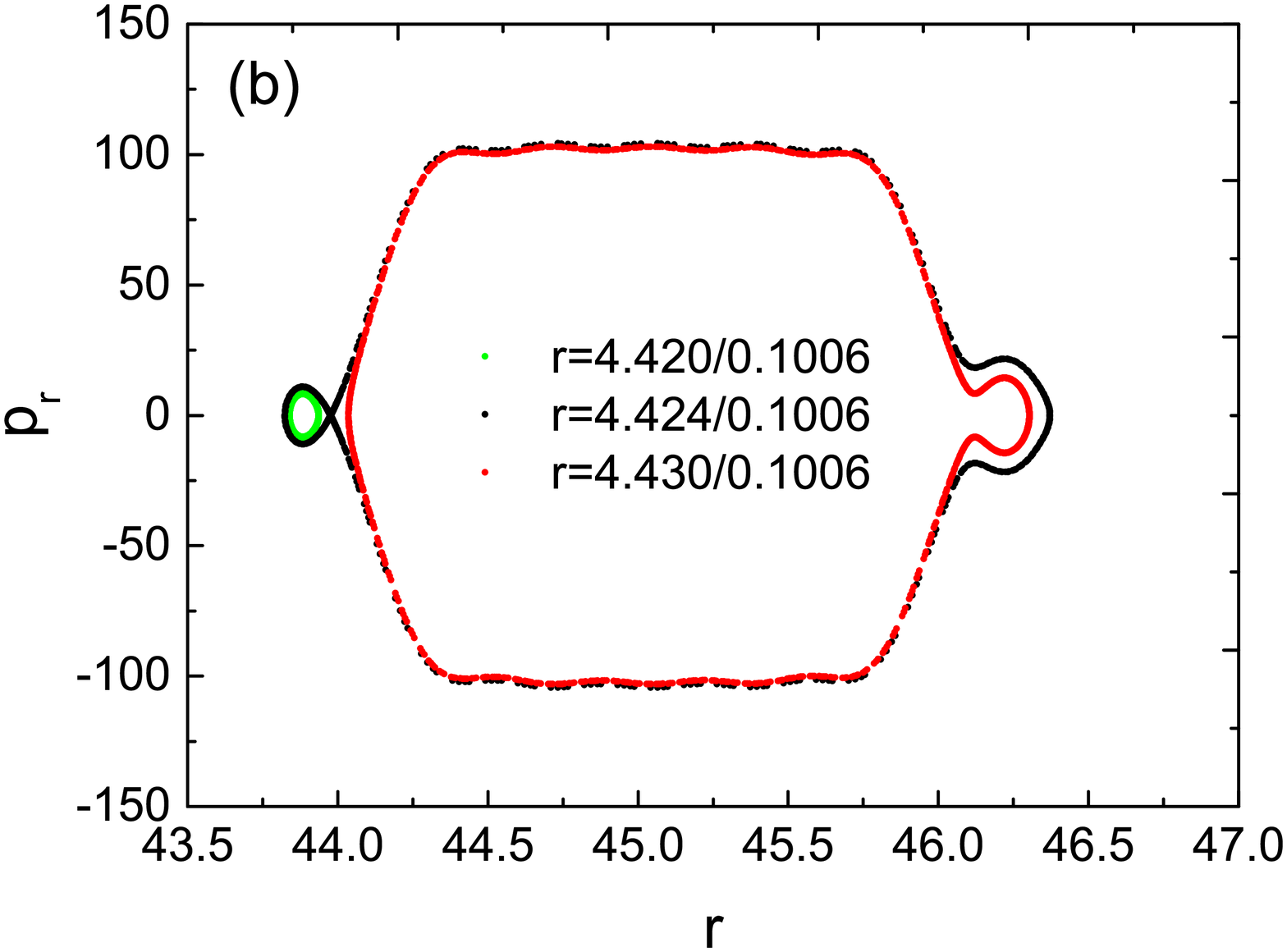}
\includegraphics[scale=0.19]{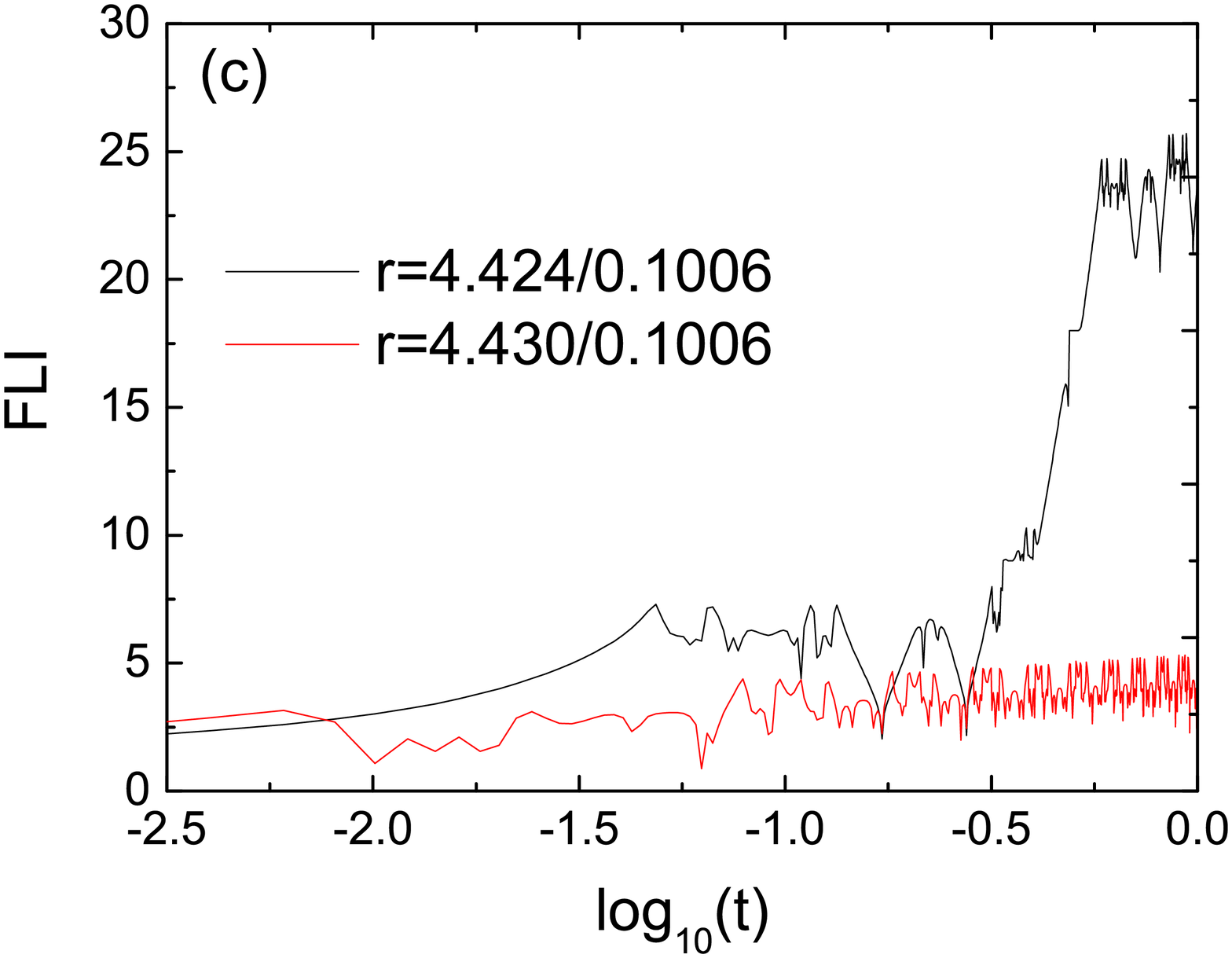}
\caption{(a) and (b): Phase space structures of Model A4 on the
Poincar\'{e} section at the  plane $\phi=\pi/2$ with $p_{\phi}>0$.
The dynamical structures of ordered and chaotic orbits have
different spatial distributions along the radial distance $r$. The
parameters and the initial conditions (except $r$ and $p_\phi$)
are those of Fig.1. The structures are considered in the range of
$10<r<12.4$ (a) and $43.5<r<46.5$ (b). (c): FLIs of the two orbits
in panel (b). The FLIs show the regularity of the orbit for
$r=4.430/0.1006$ and the chaoticity of the orbit for
$r=4.424/0.1006$. }} \label{f2}
\end{figure*}

\begin{figure*}
\center{
\includegraphics[scale=0.19]{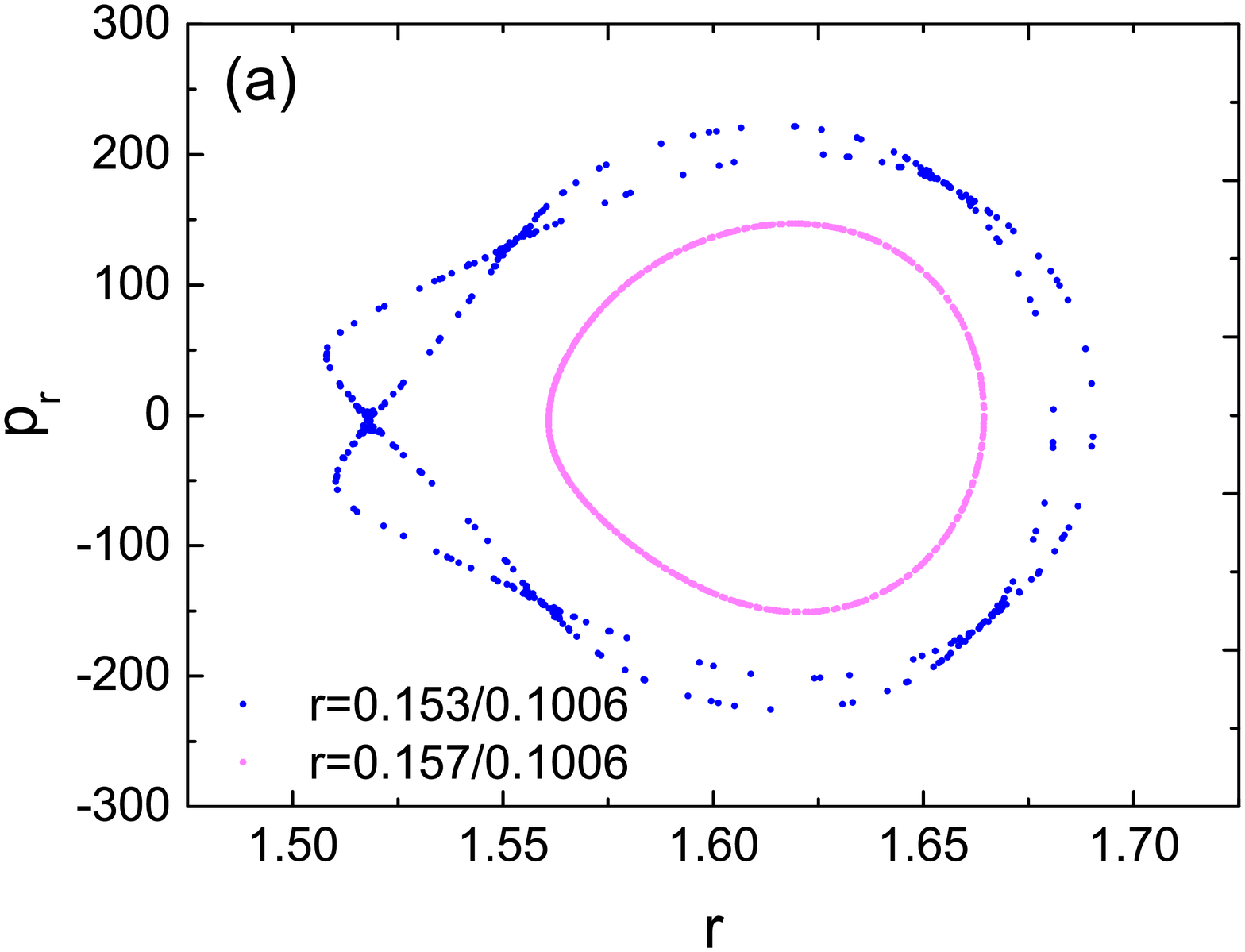}
\includegraphics[scale=0.19]{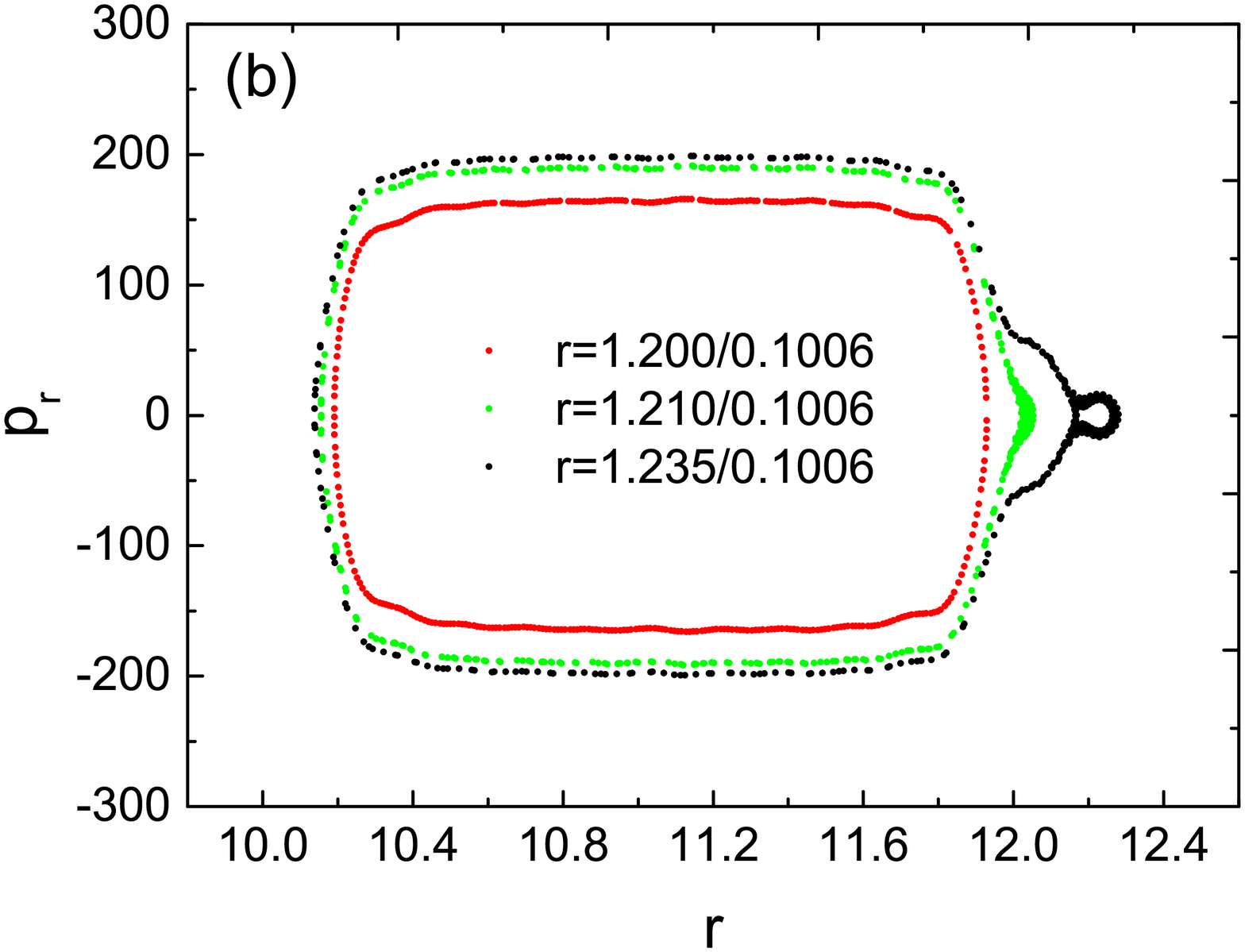}
\includegraphics[scale=0.19]{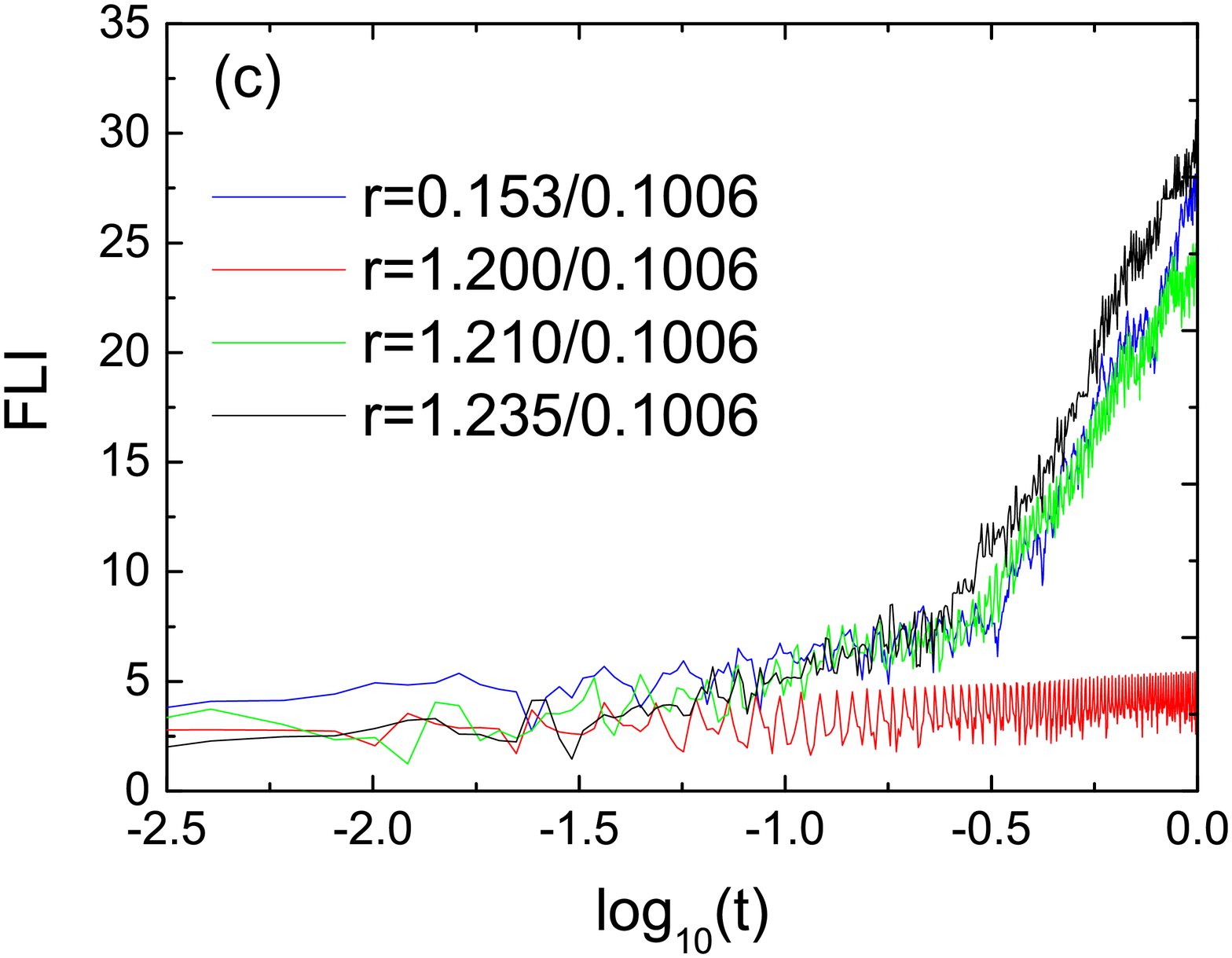}
\caption{(a) and (b): Phase space structures of Model A9 on the
Poincar\'{e} section. The structures are considered in the ranges
of $1.5<r<1.7$ (a) and $10<r<12.4$ (b). The initial conditions in
Fig. 3(b) are the same as those in Fig. 2(a). However, the orbits
for $r=1.210/0.1006$ and $r=1.235/0.1006$ are chaotic, whereas the
orbit for $r=1.200/0.1006$ is regular. (c): FLIs of several orbits
in panels (a) and (b). The FLIs show the regularity of the orbit
for $r=1.200/0.1006$ and the chaoticity of the three orbits for
$r=0.153/0.1006$, $r=1.210/0.1006$ and $r=1.235/0.1006$. }}
\label{f3}
\end{figure*}

\begin{figure*}
\center{
\includegraphics[scale=0.26]{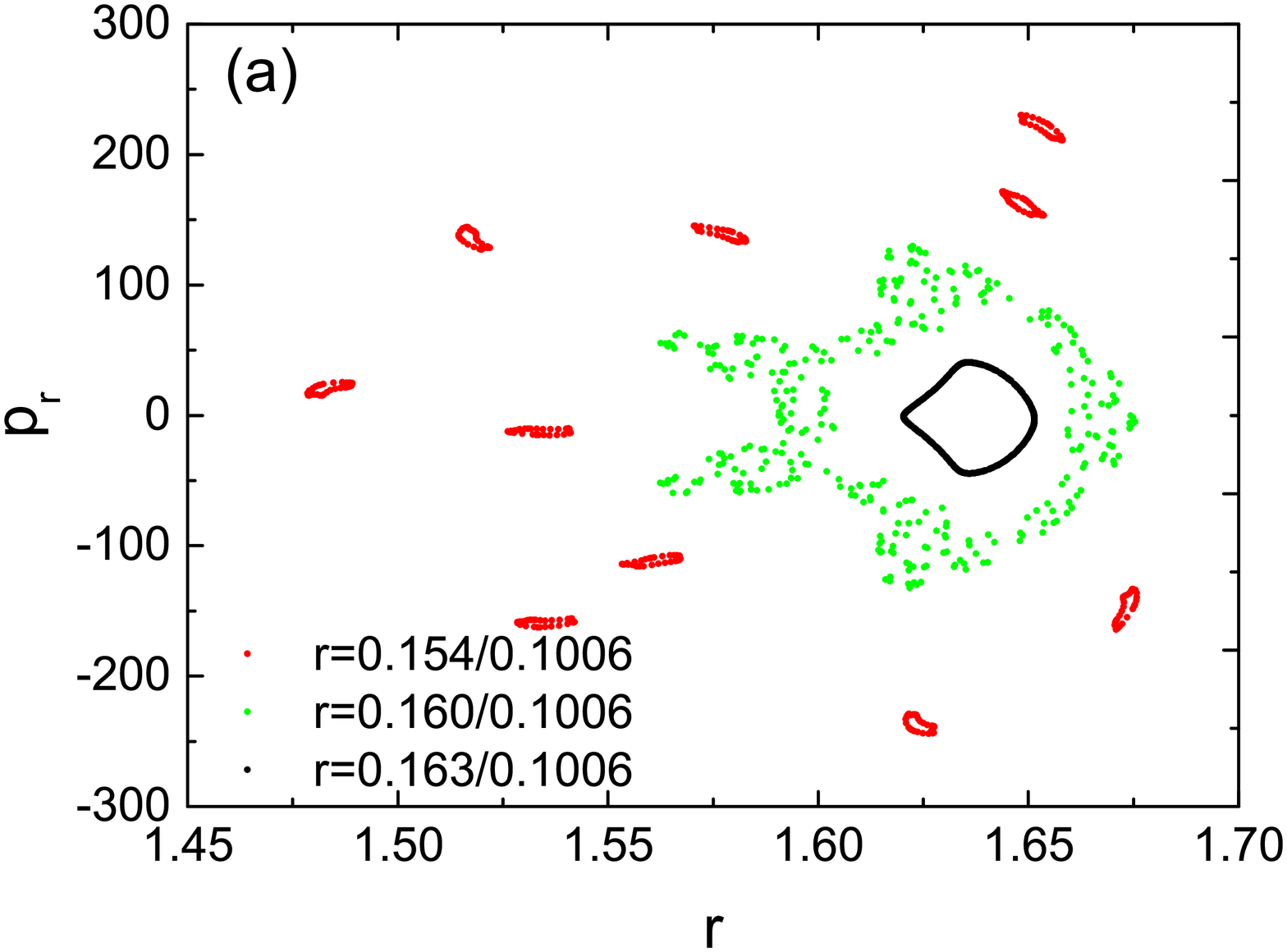}
\includegraphics[scale=0.26]{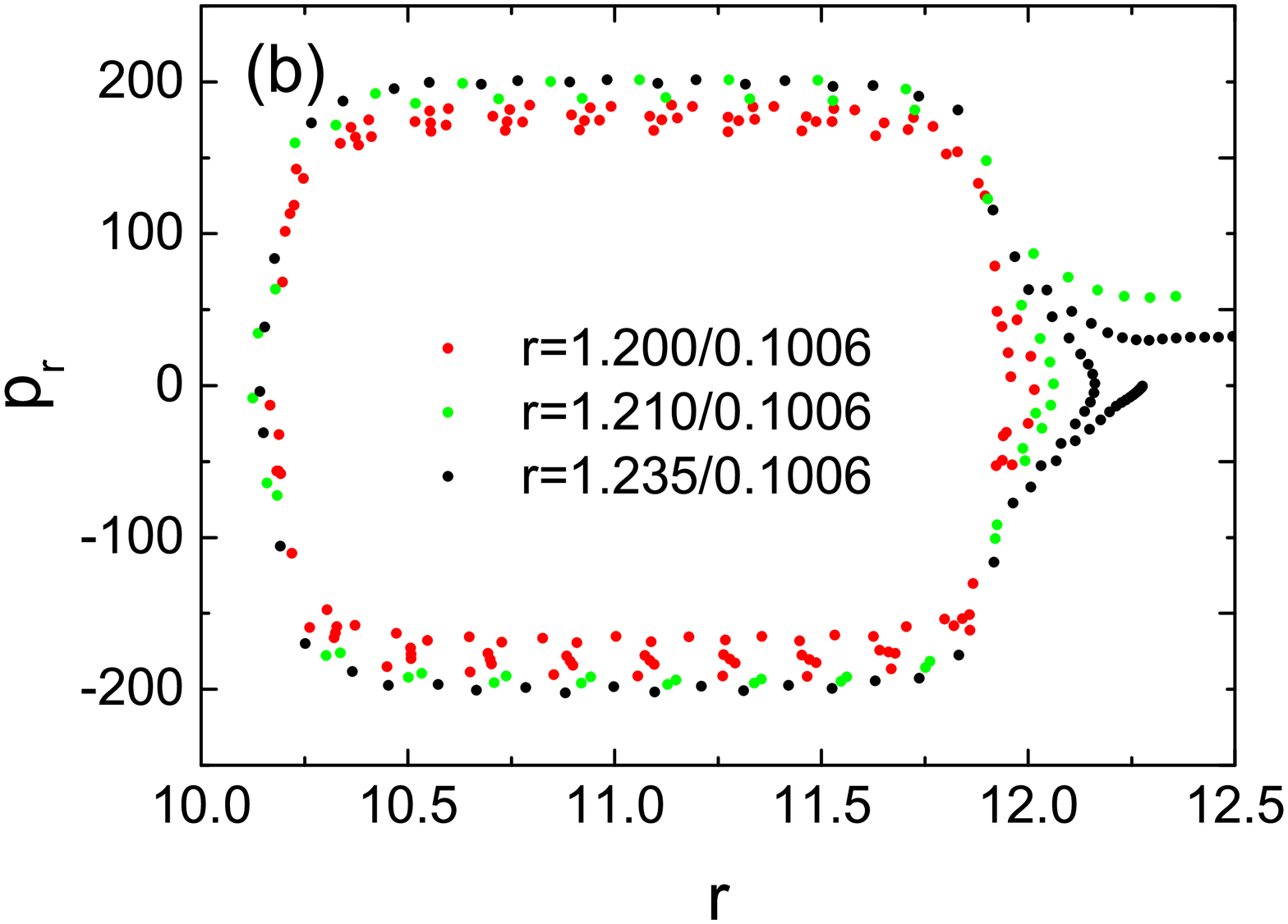}
\includegraphics[scale=0.26]{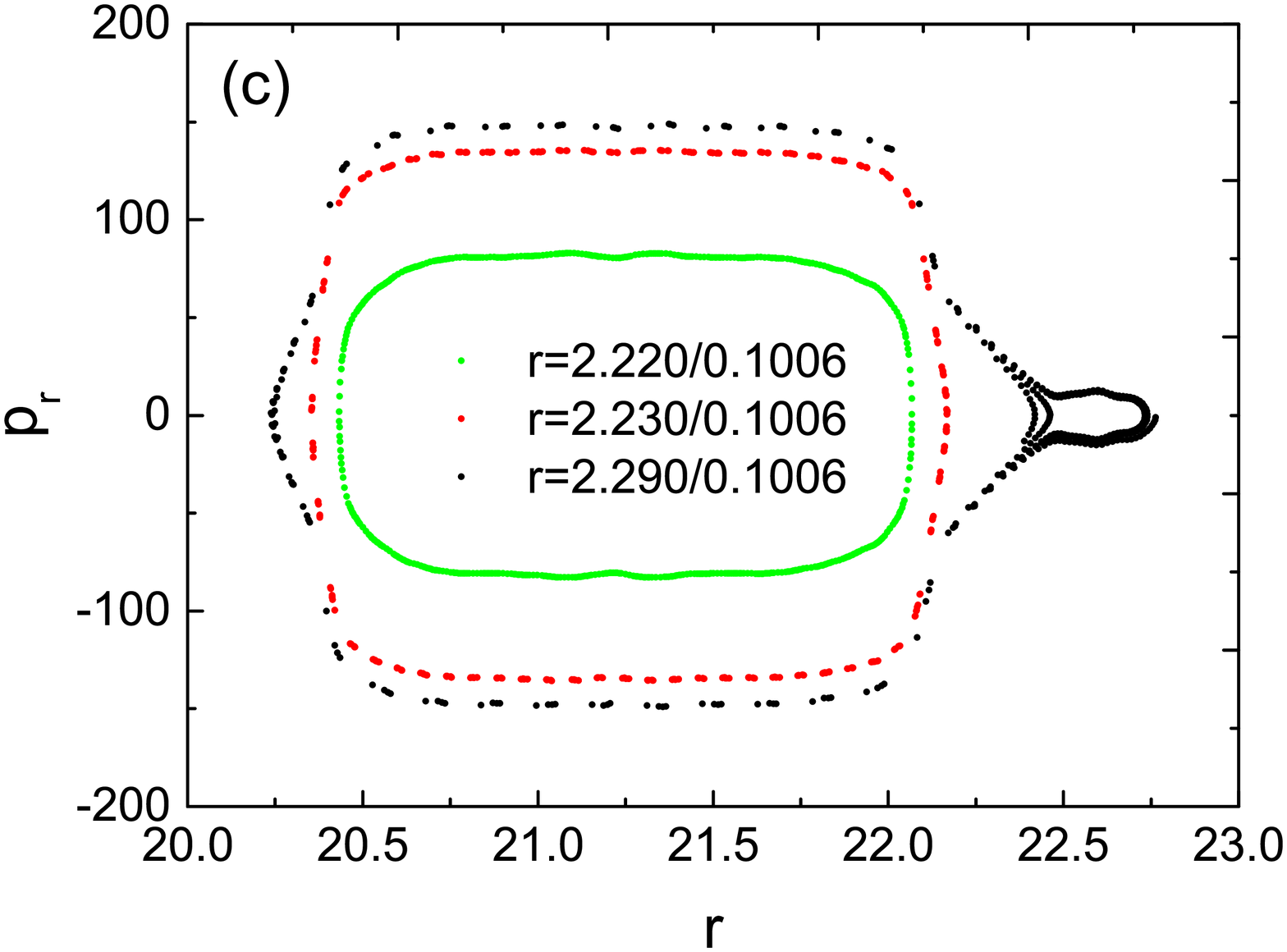}
\includegraphics[scale=0.26]{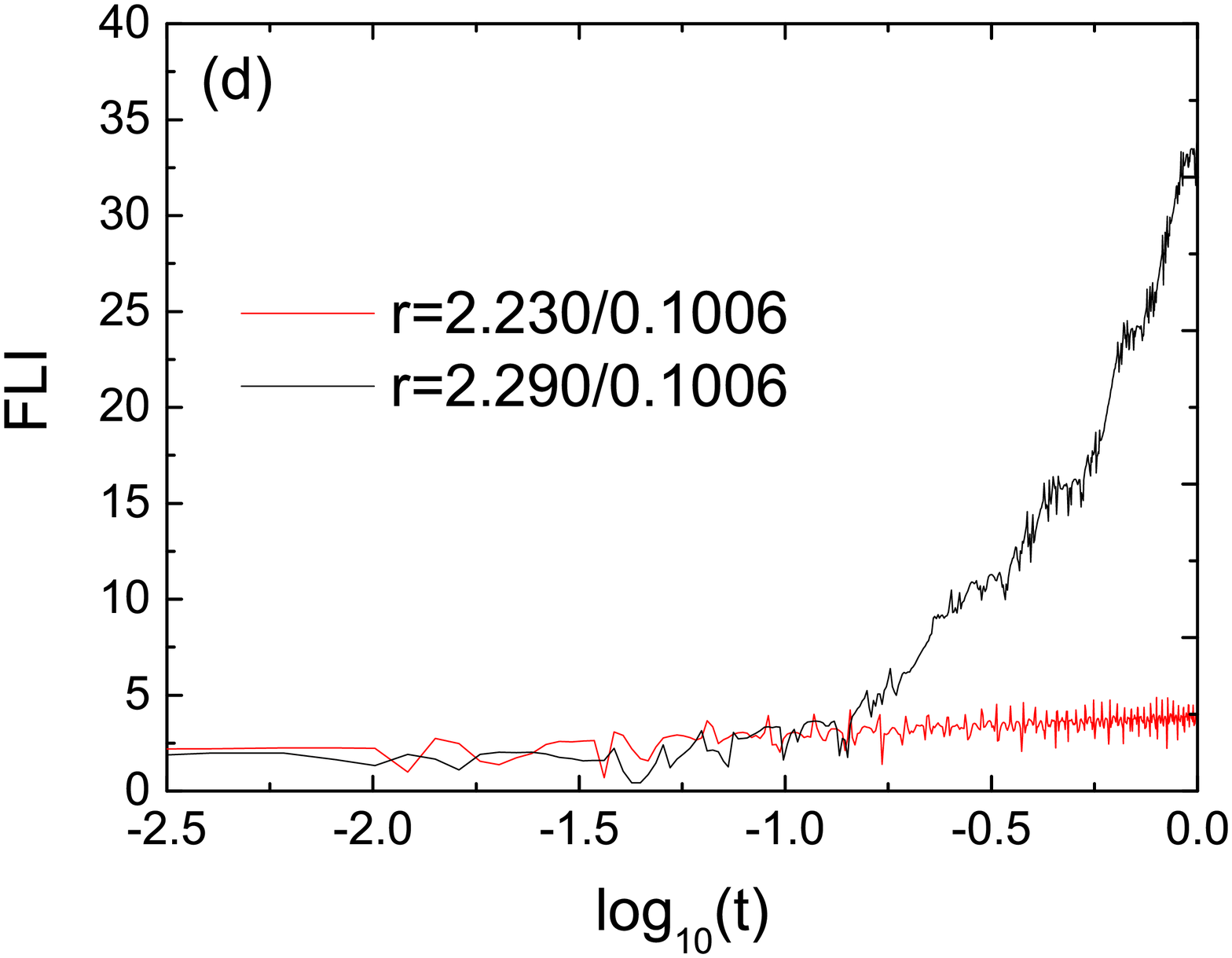}
\caption{(a)-(c): Phase space structures of Model A19 on the
Poincar\'{e} section. (a): The structures are considered in the
range of $1.45<r<1.7$. The orbit for $r=0.163/0.1006$ is an
ordered KAM torus, and the orbit for $r=0.154/0.1006$ with many
islands is still regular. However, the orbit for $r=0.160/0.1006$
is chaotic. (b): The structures are considered in the range of
$10<r<12.4$. The initial conditions in Fig. 4(b) are the same as
those in Fig. 2(a). (c): The structures are considered in the
range of $20<r<23$. There are a number of regular KAM torus orbits
in the interior of the ordered orbit for $r=2.230/0.1006$. A small
chaotic region between the two orbits for $r=2.230/0.1006$ and
$r=2.290/0.1006$ exists. (d): Fast Lyapunov indicators (FLIs) of
the two orbits in panel (c). The orbit for $r=2.230/0.1006$ having
a power law increase of the FLI with time $\log_{10}t$ is ordered,
but the orbit for $r=2.290/0.1006$ having an exponential law
increase of the FLI with time is chaotic. The FLIs are calculated
until the number of integration steps reaches $5\times 10^{4}$. }}
\label{f4}
\end{figure*}

\begin{figure*}
\center{
\includegraphics[scale=0.20]{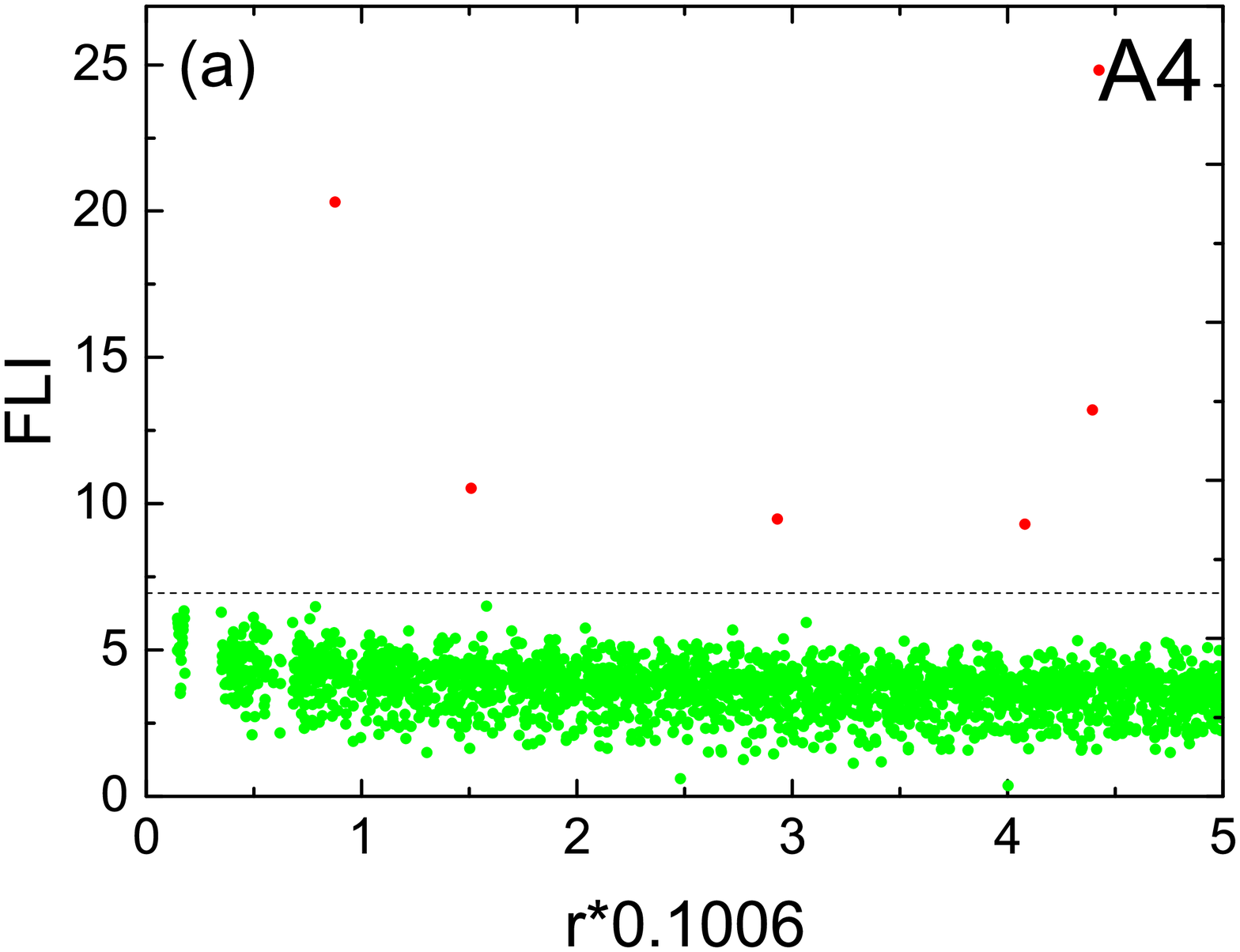}
\includegraphics[scale=0.20]{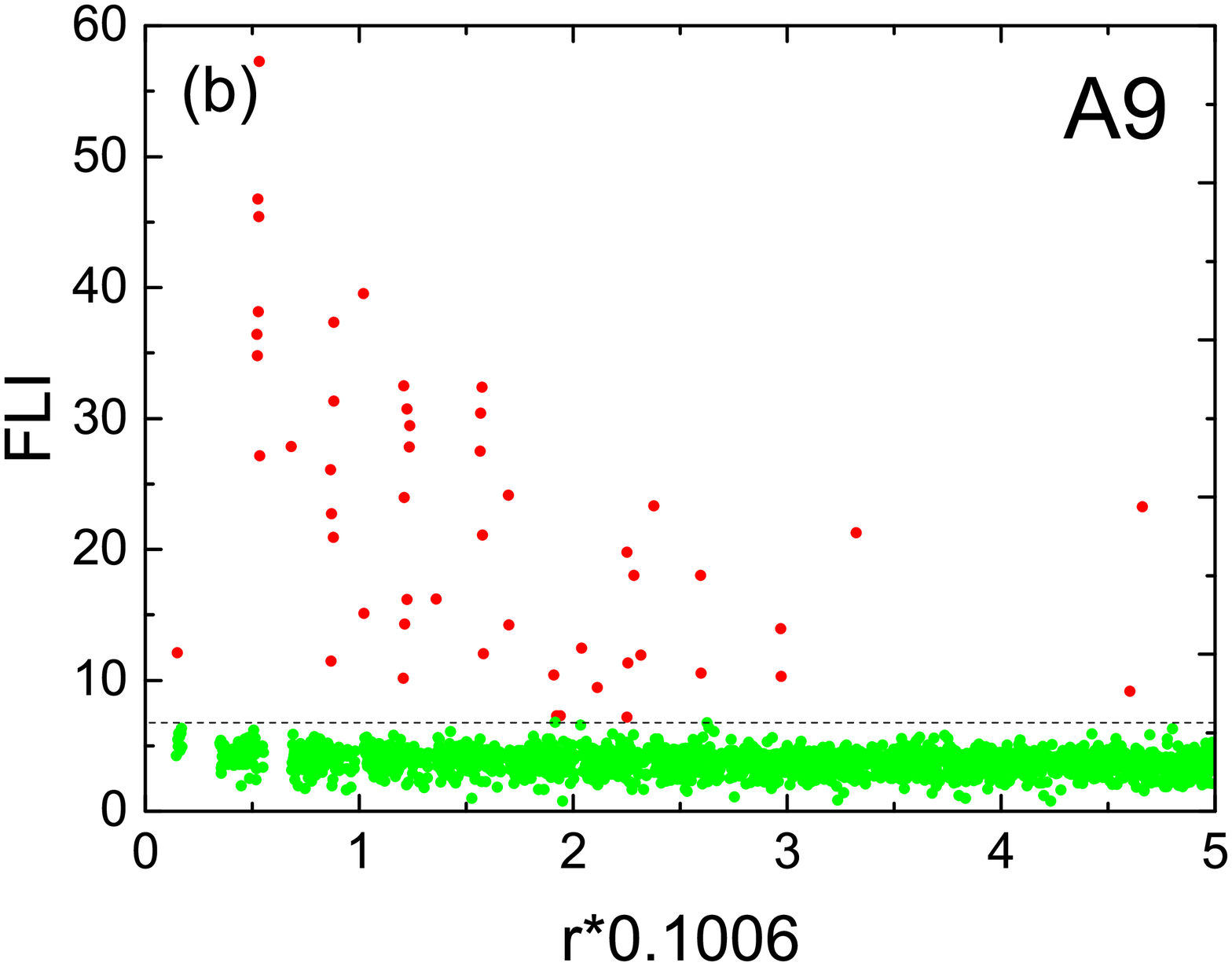}
\includegraphics[scale=0.20]{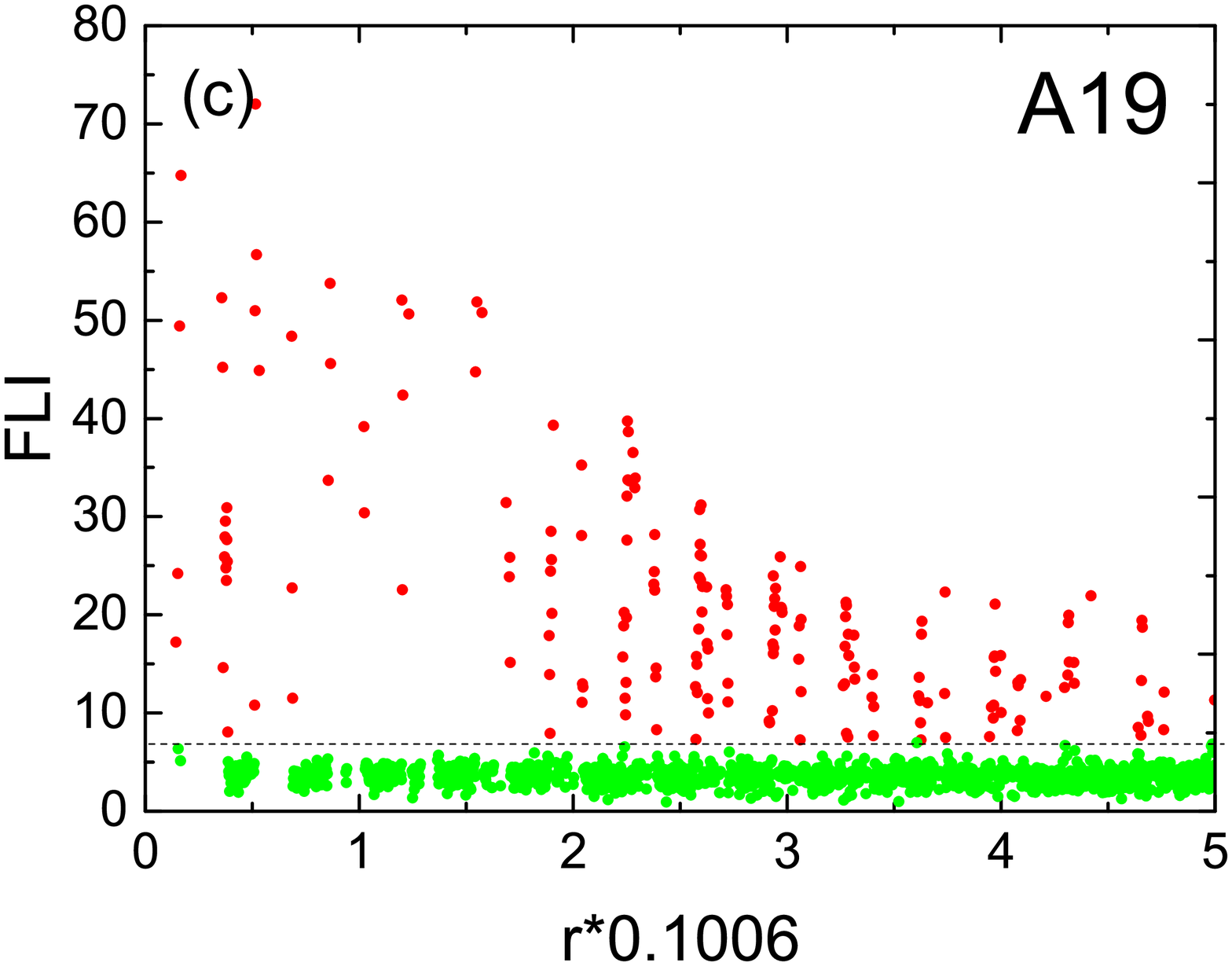}
\caption{The dependence of FLI on the initial separation $r$ in
Models (a) A4, (b) A9 and (c) A19. Each of the FLIs is obtained
after $5\times10^4$ integration steps. The FLIs not more  than 7
show the regularity of bounded orbits, whereas those larger than 7
indicate the chaoticity of bounded orbits. Many values  of $r$
correspond to chaos in Model A19, but only minor values  of $r$ do
in Model A4. It is clear that chaos becomes easier and its degree
is enhanced as the radial term number $n$ increases.}} \label{f5}
\end{figure*}

\begin{figure*}
\center{
\includegraphics[scale=0.28]{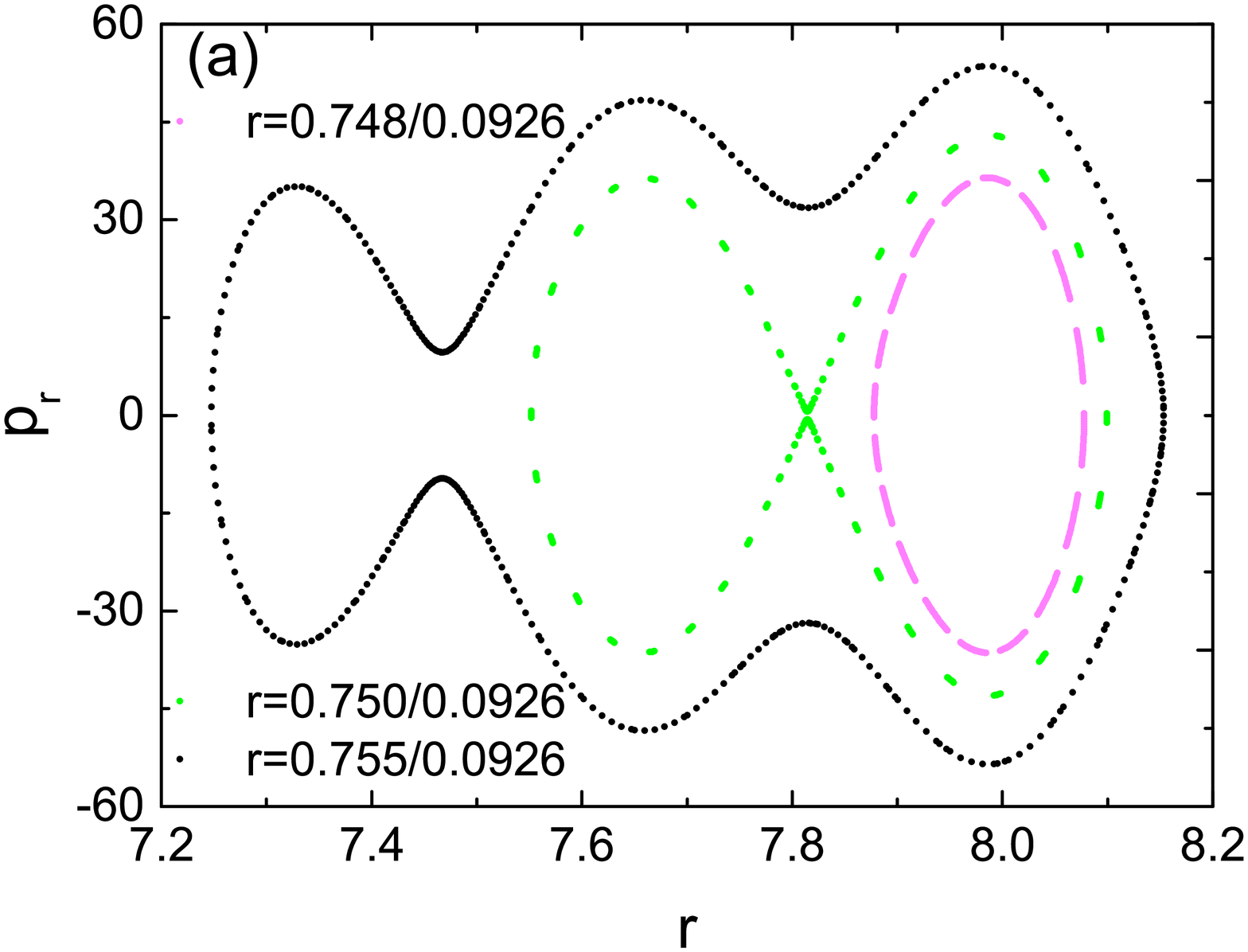}
\includegraphics[scale=0.28]{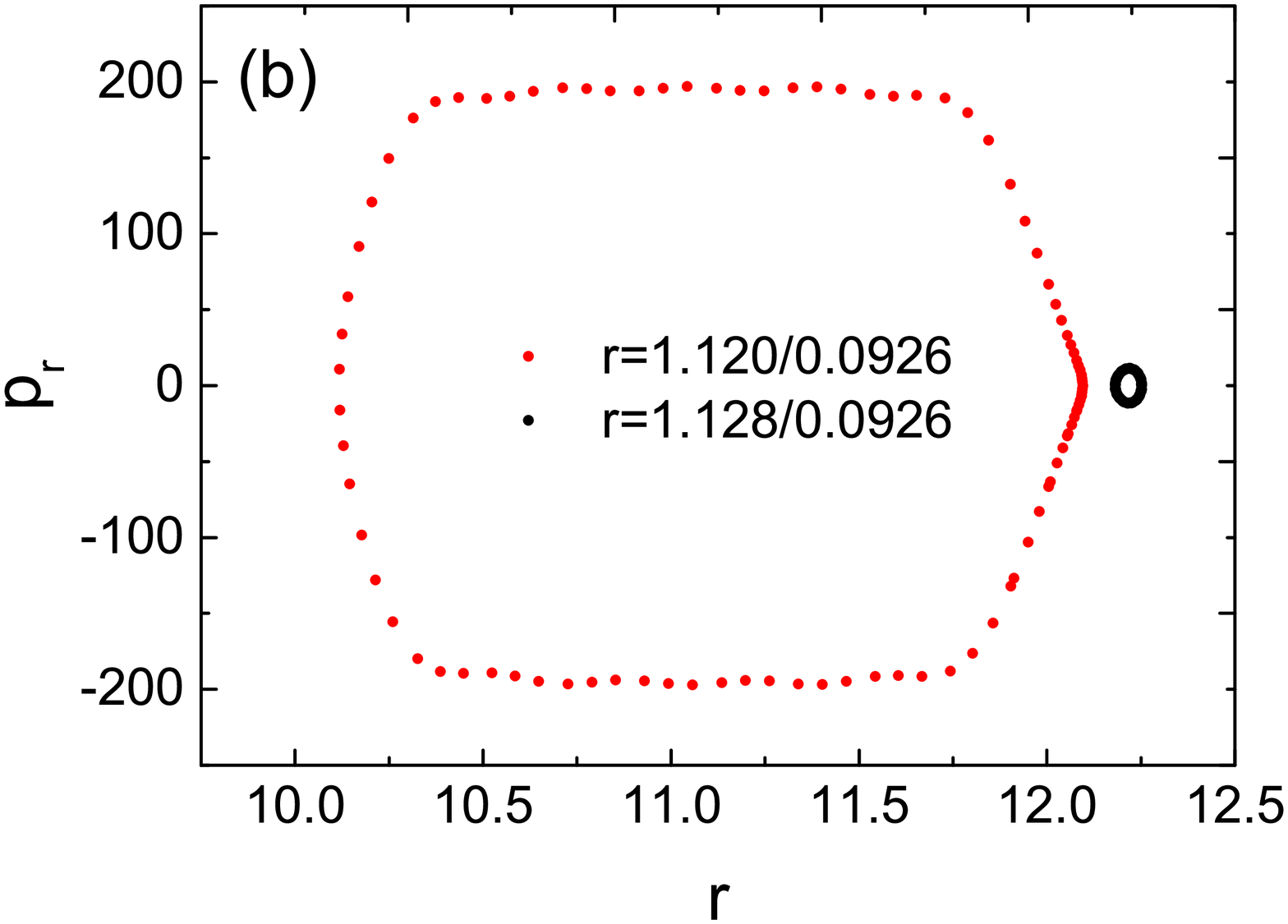}
\includegraphics[scale=0.28]{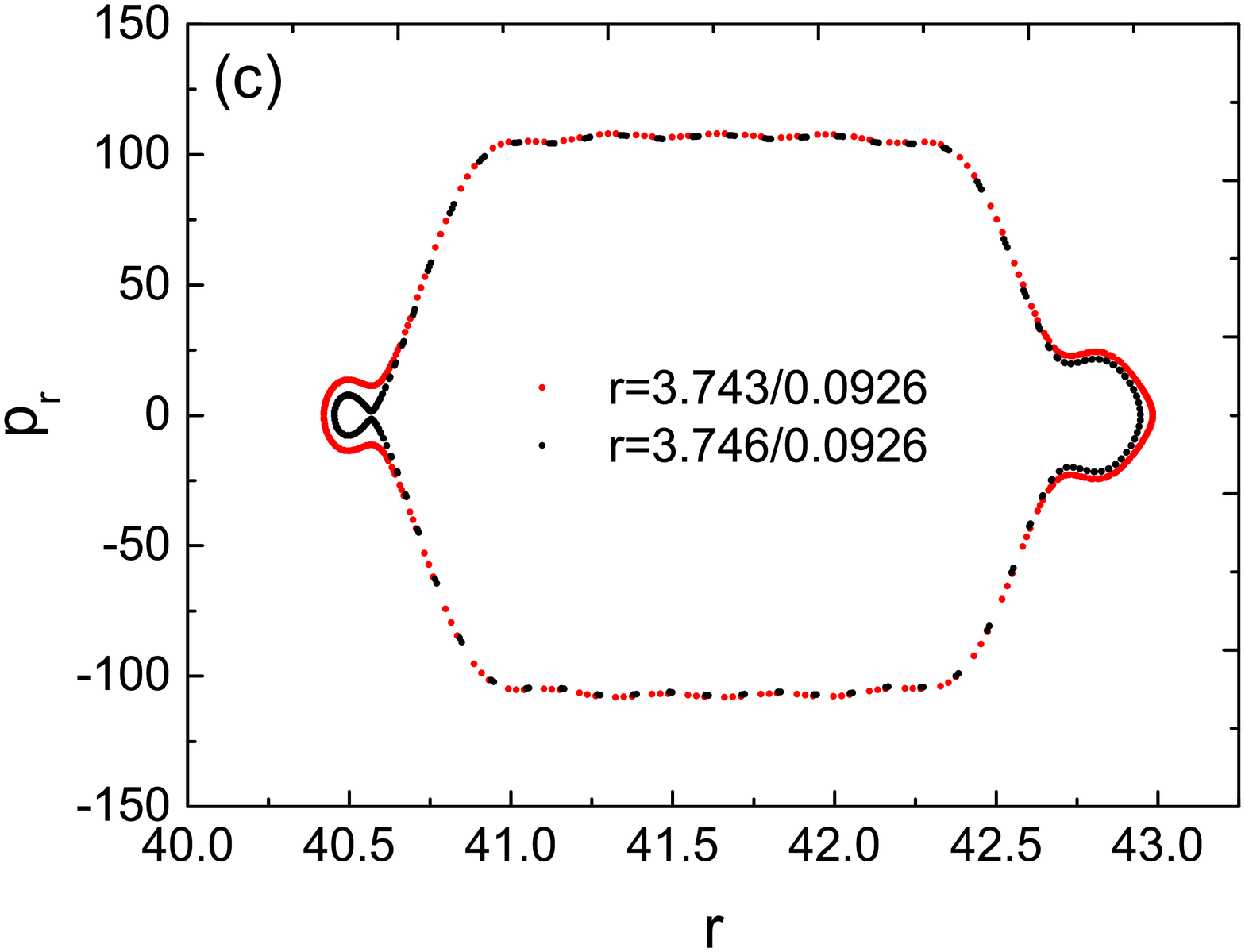}
\includegraphics[scale=0.28]{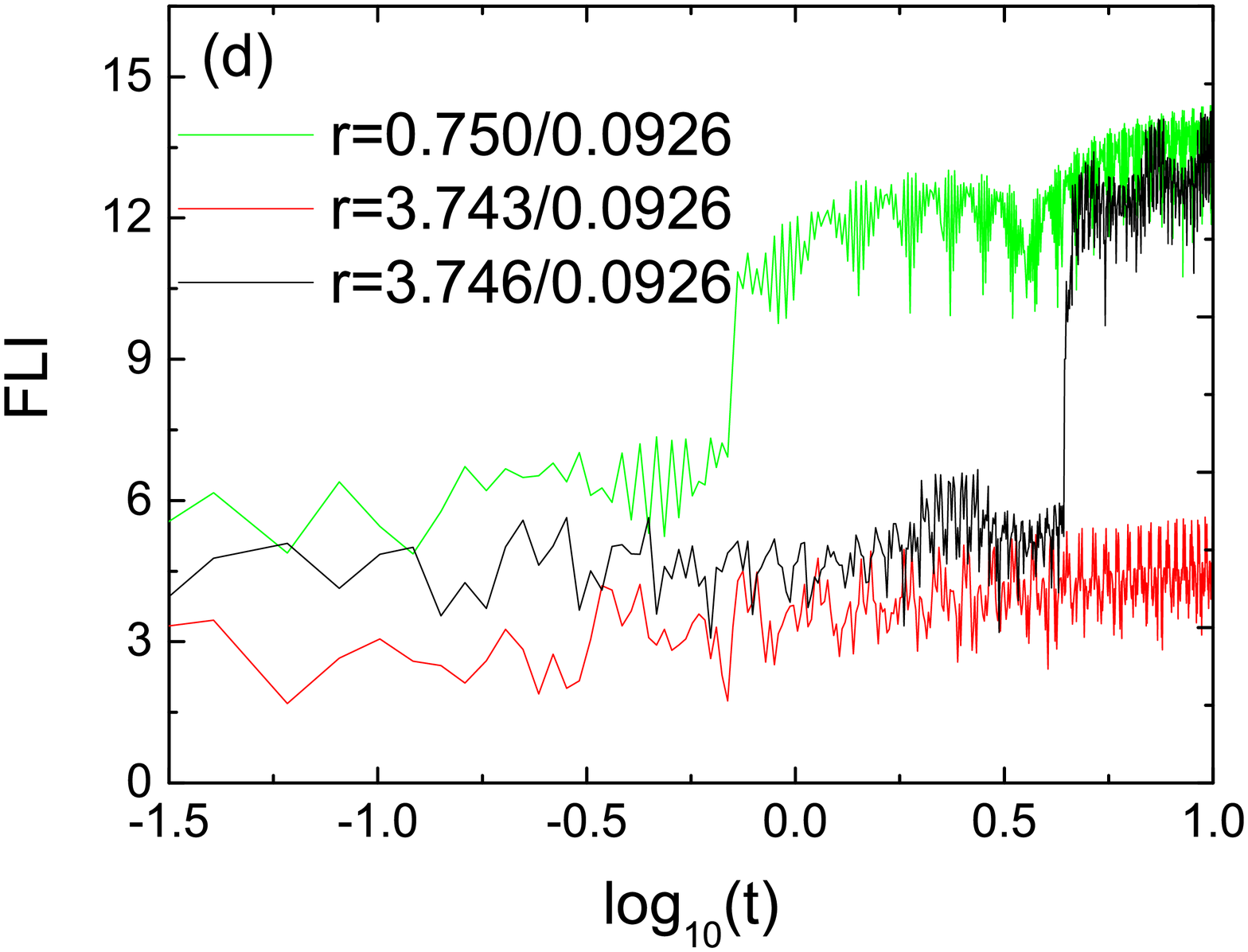}
\caption{Similar to Fig. 2 but Model B4 is considered. (a):  The
structures are considered in the range of $7.2<r<8.2$. (b):  The
structures are considered in the range of $12.0<r<12.2$. (c):  The
structures are considered in the range of $40<r<43$ and resemble
those in Fig. 2(b). (d): The FLIs show the chaoticity of the two
orbits in panels (a) and (c) and the regularity of the orbit in
panel (c). }} \label{f6}
\end{figure*}

\begin{figure*}
\center{
\includegraphics[scale=0.185]{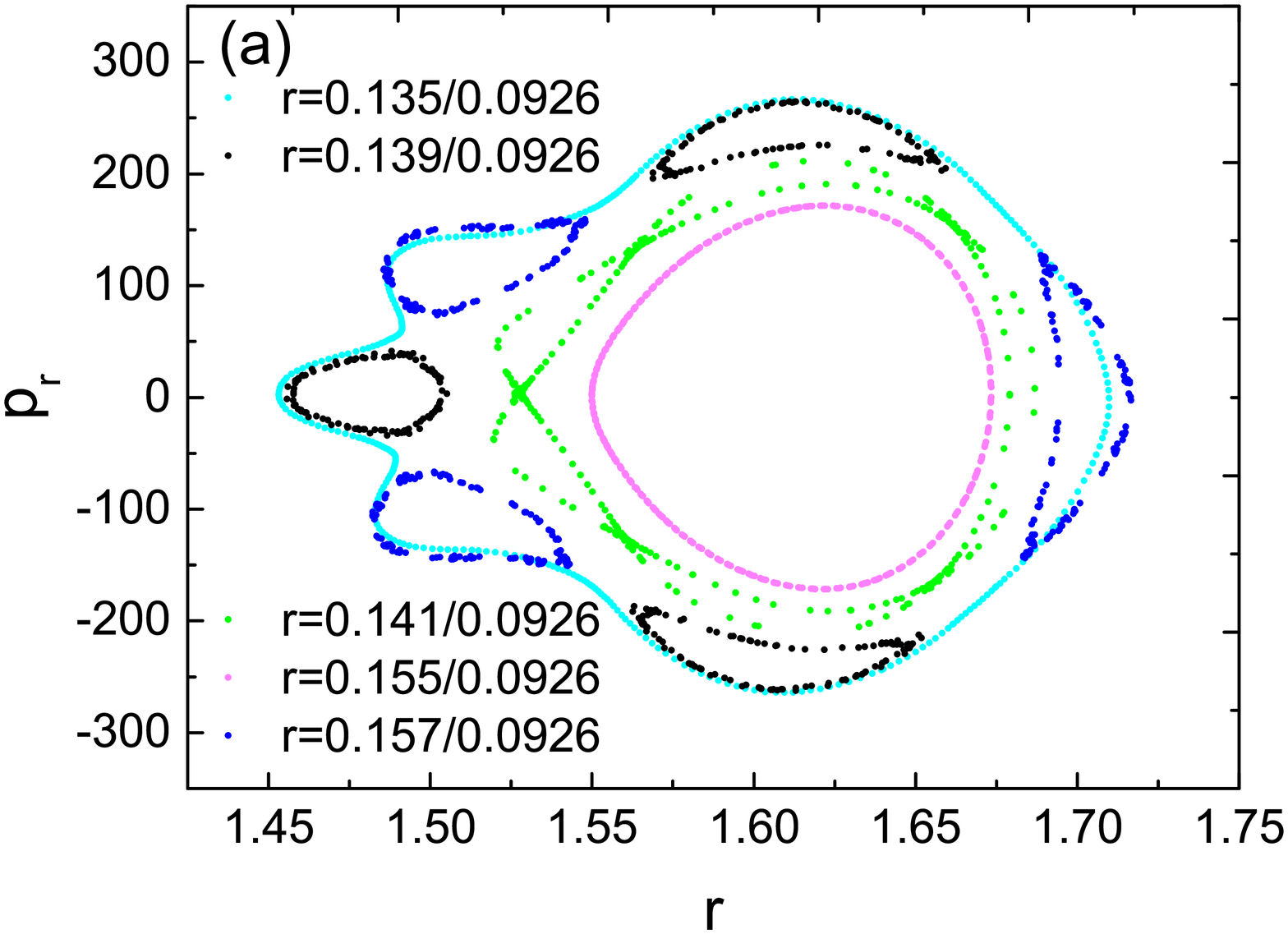}
\includegraphics[scale=0.185]{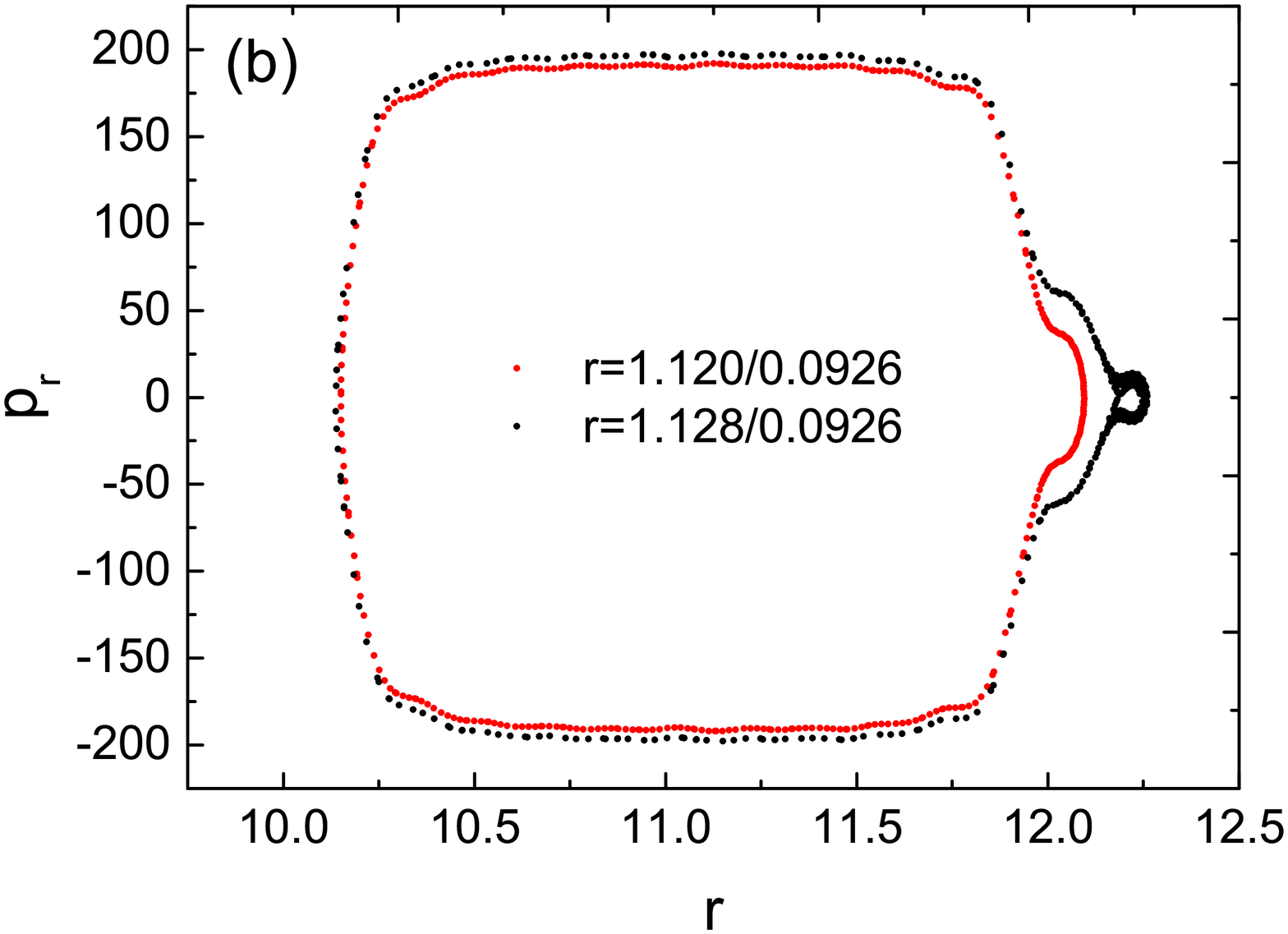}
\includegraphics[scale=0.185]{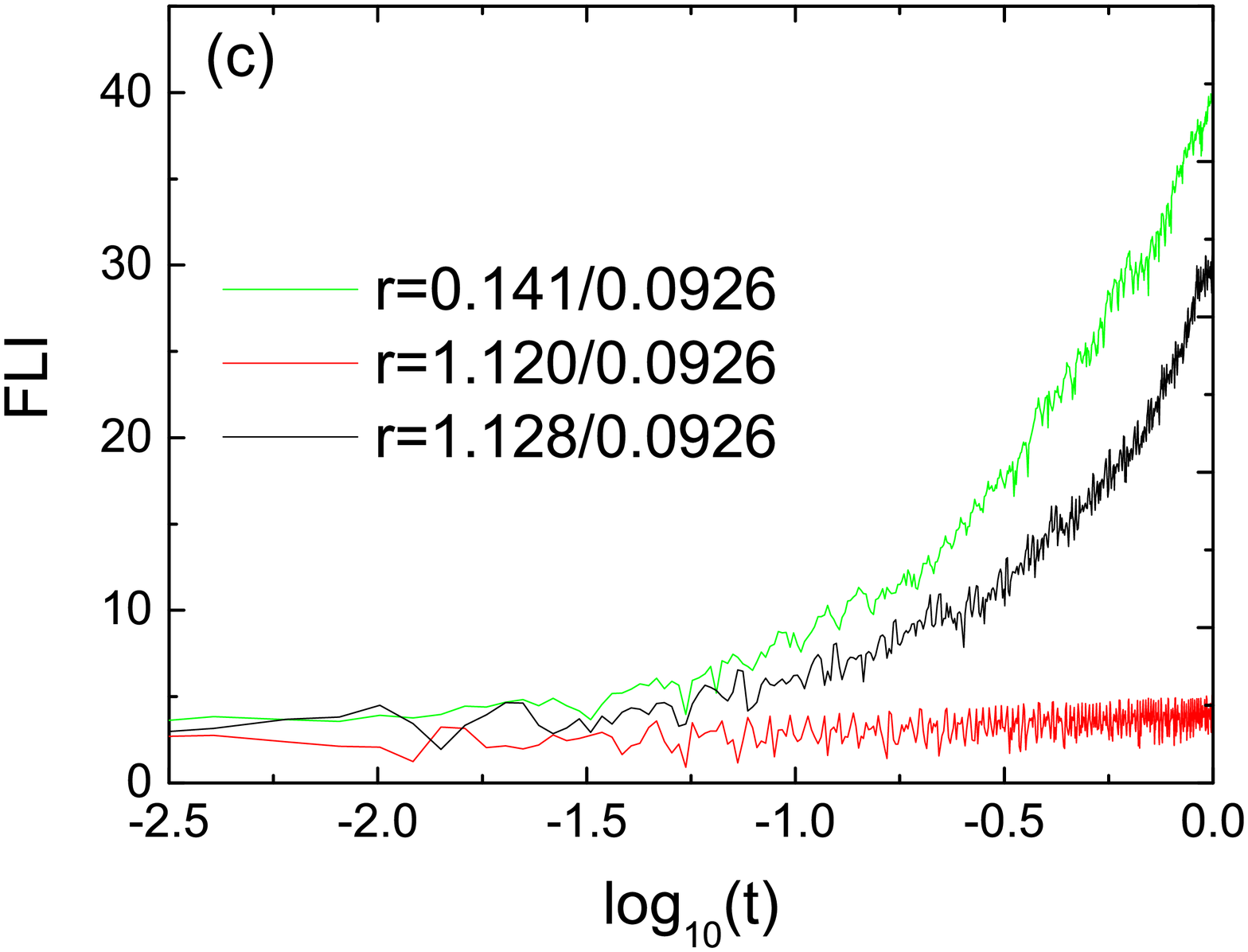}
\caption{Similar to Fig. 3 but Model B9 is considered. The phase
space structures in panels (a) and (b) respectively look like
those in Fig. 3 (a) and (b). The initial conditions in Fig. 7(b)
are the same as those in Fig. 6(b). The orbit for $r=1.128/0.0926$
is chaotic, but the orbit for $r=1.120/0.0926$ is regular. (c):
The FLIs show the chaoticity of the two orbits in panels (a) and
(b) and the regularity of the orbit in panel (b). }} \label{f7}
\end{figure*}

\begin{figure*}
\center{
\includegraphics[scale=0.26]{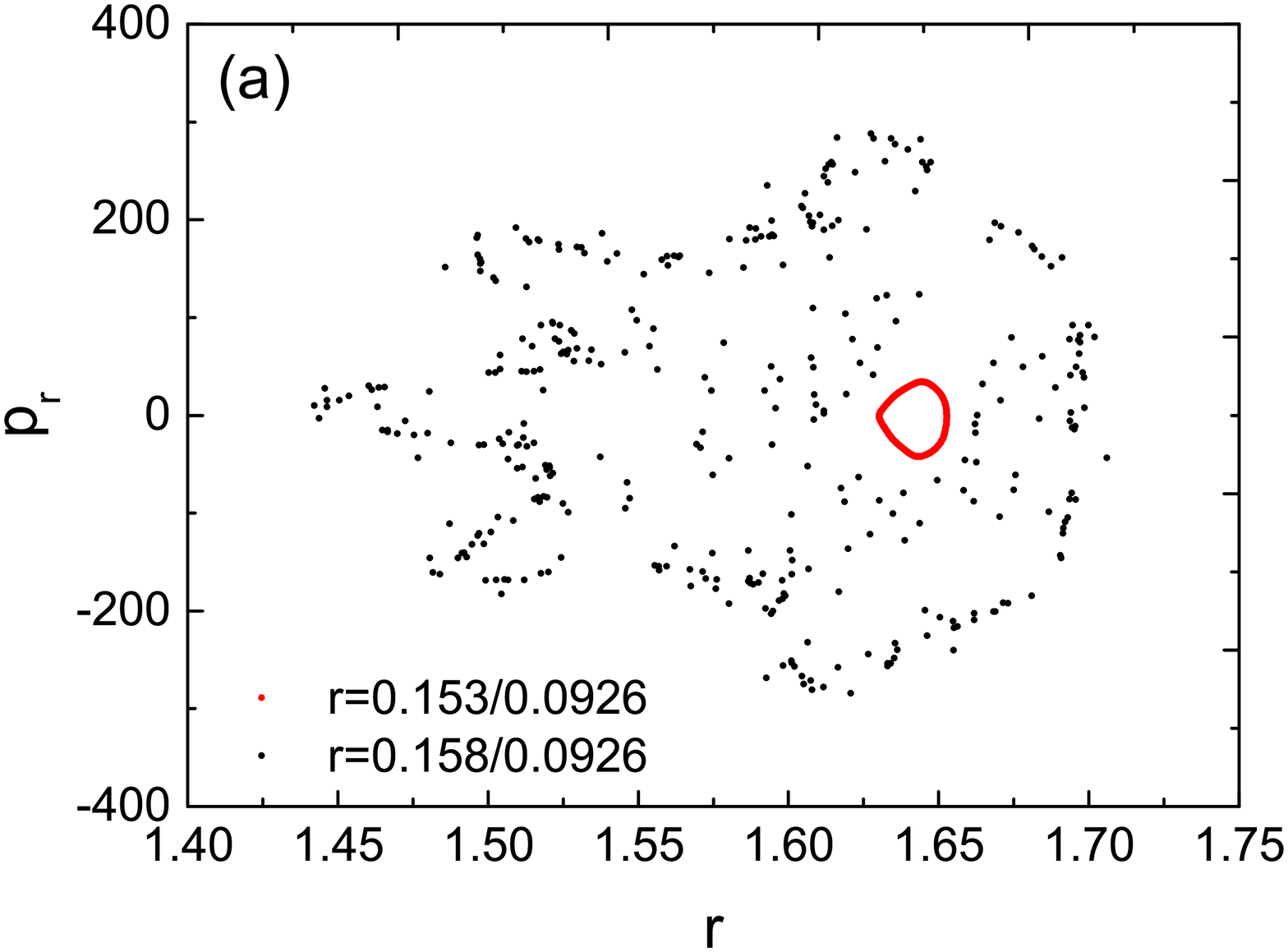}
\includegraphics[scale=0.26]{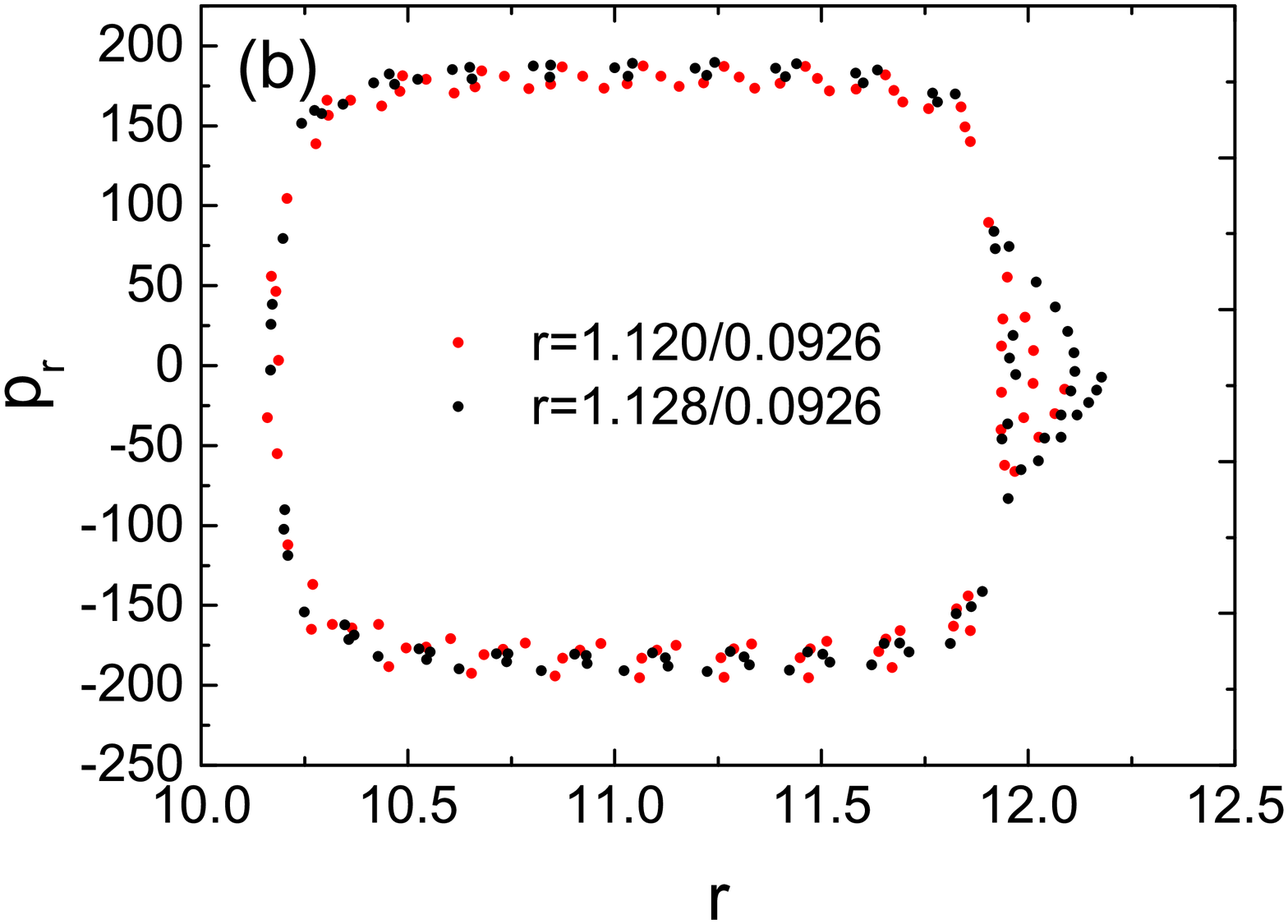}
\includegraphics[scale=0.26]{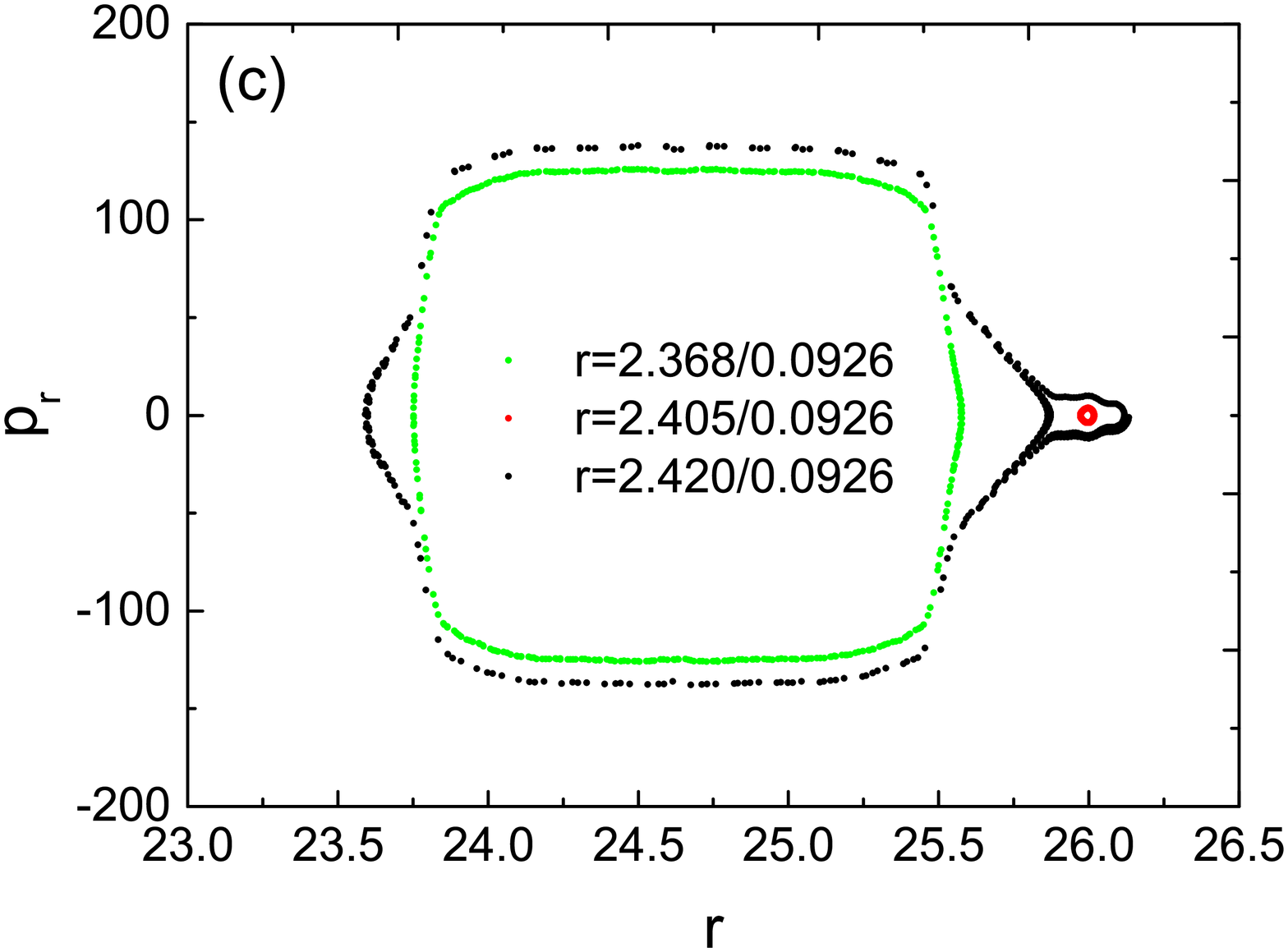}
\includegraphics[scale=0.26]{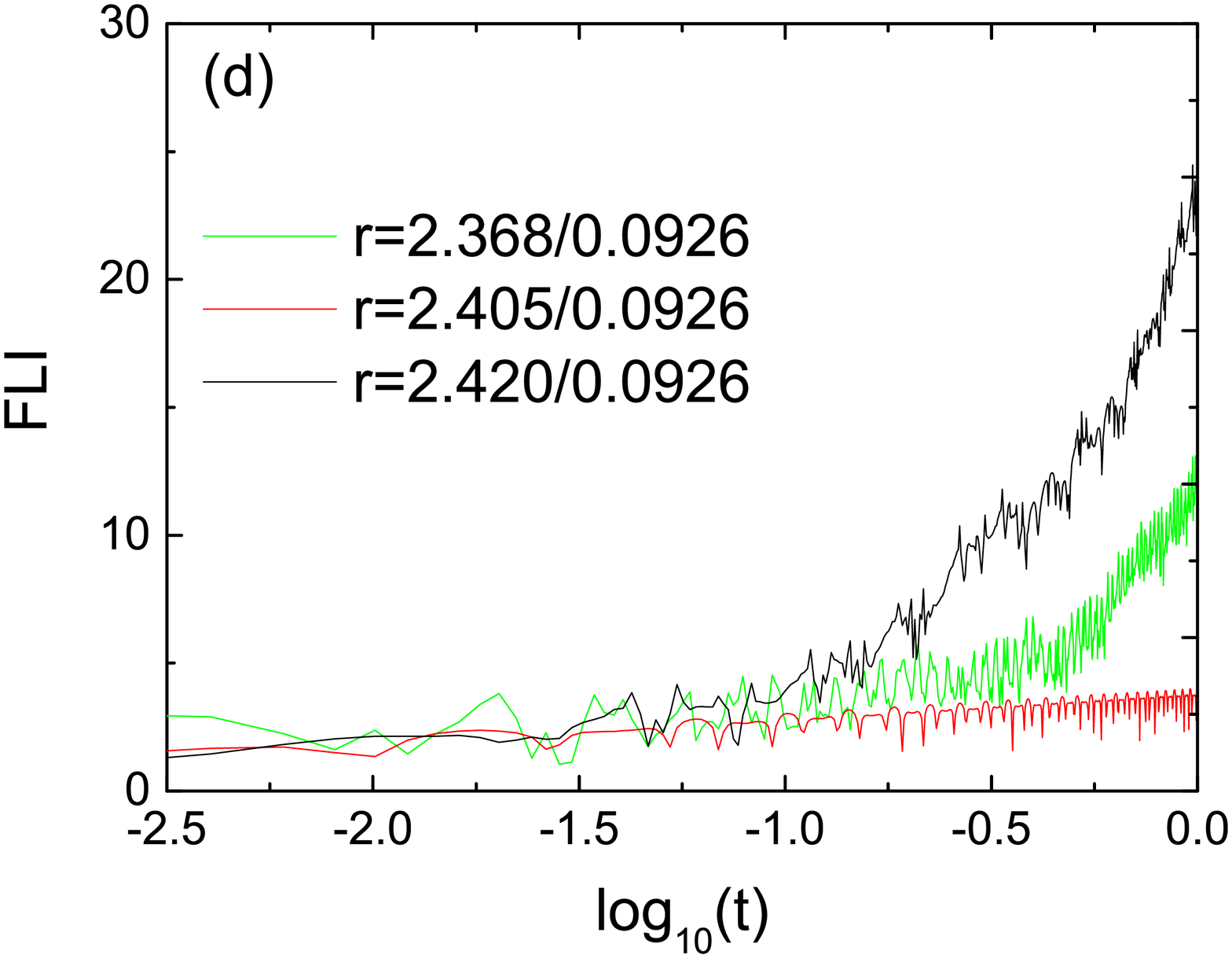}
\caption{Similar to Fig. 4 but Model B19 is considered. The phase
space structures in panel (a) are somewhat different from those in
Fig. 4(a), but these structures in panels (b) and (c) respectively
resemble those in Fig. 2 (b) and (c). The initial conditions in
Fig. 8(b) are the same as those in Fig. 6(b). (d): The FLIs show
the chaoticity of the two orbits in panel (c) and the regularity
of the orbit in panel (c). }} \label{f8}
\end{figure*}

\begin{figure*}
\center{
\includegraphics[scale=0.20]{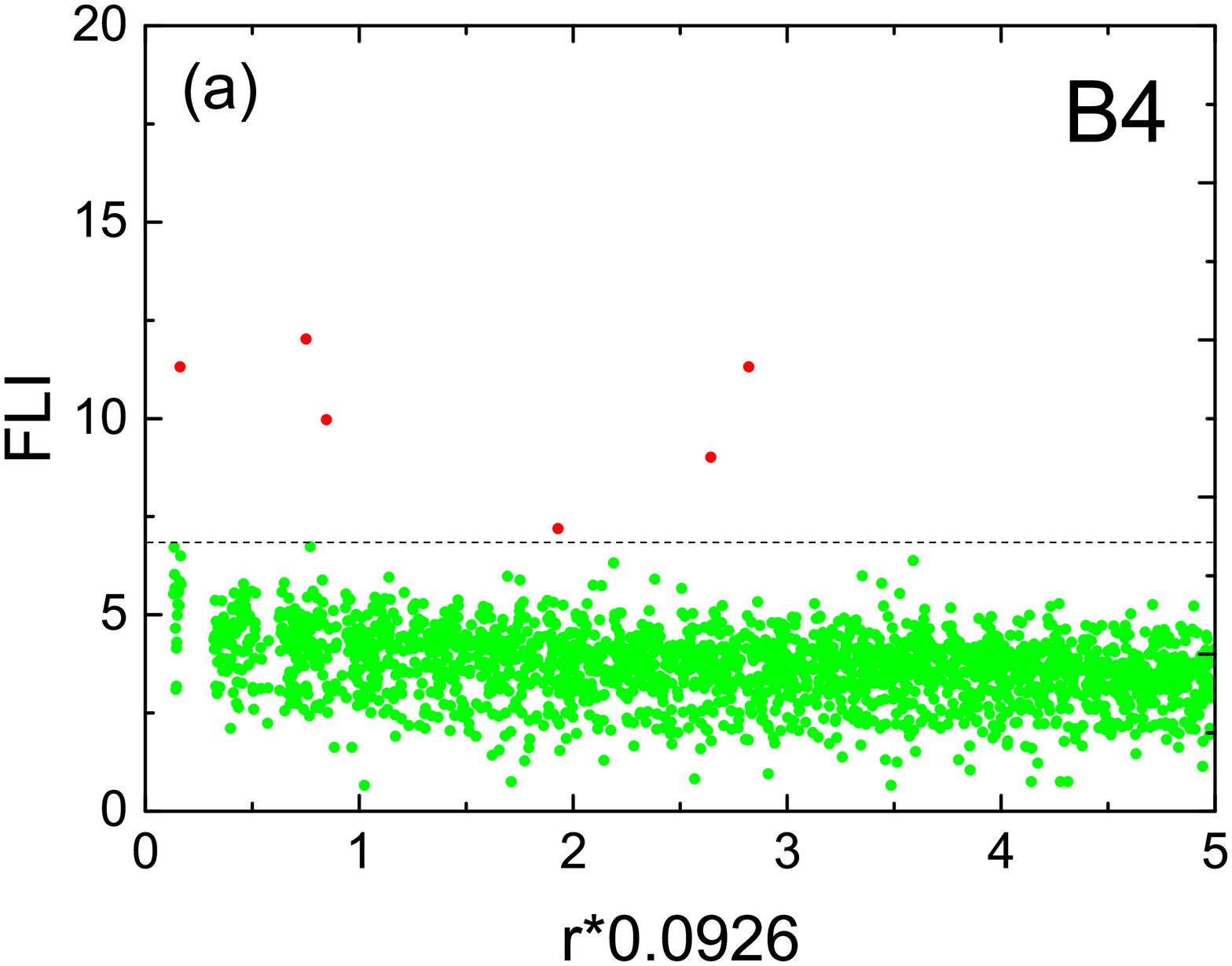}
\includegraphics[scale=0.20]{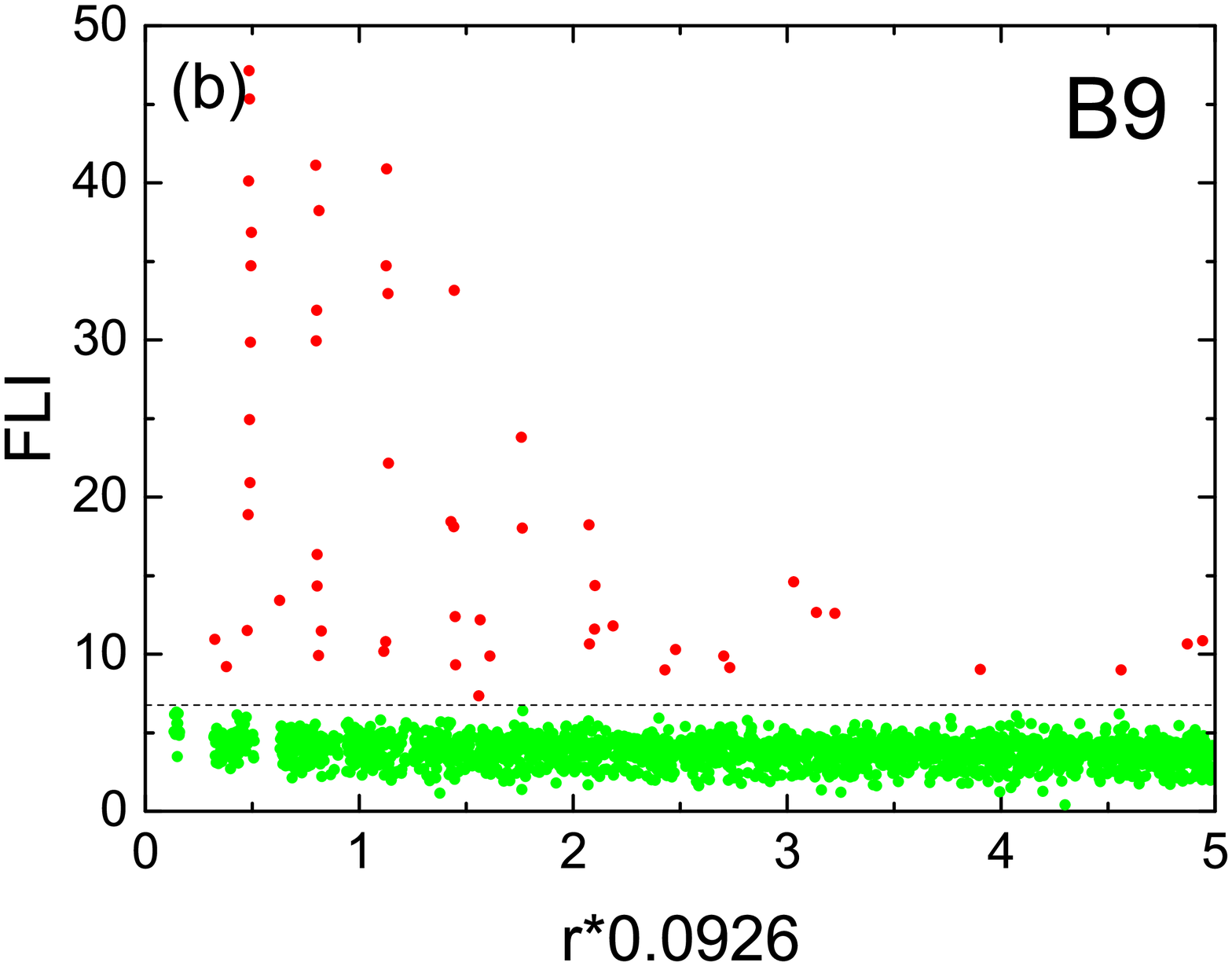}
\includegraphics[scale=0.20]{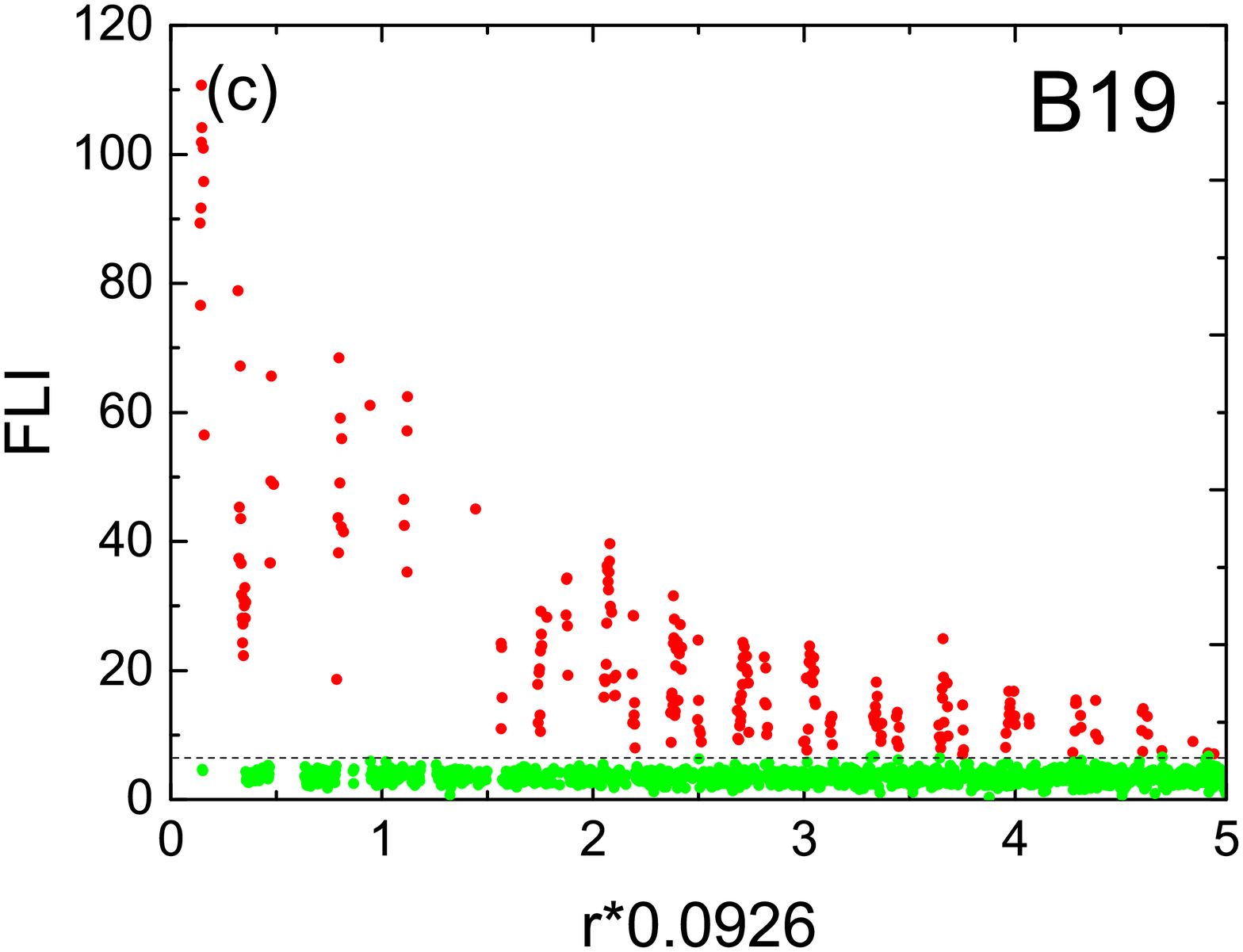}
\caption{Similar to Fig. 5 but Models A are replaced with Models
B. The degree of chaos becomes stronger with an increase of the
radial term number $n$.}} \label{f9}
\end{figure*}

\begin{figure*}
\center{
\includegraphics[scale=0.20]{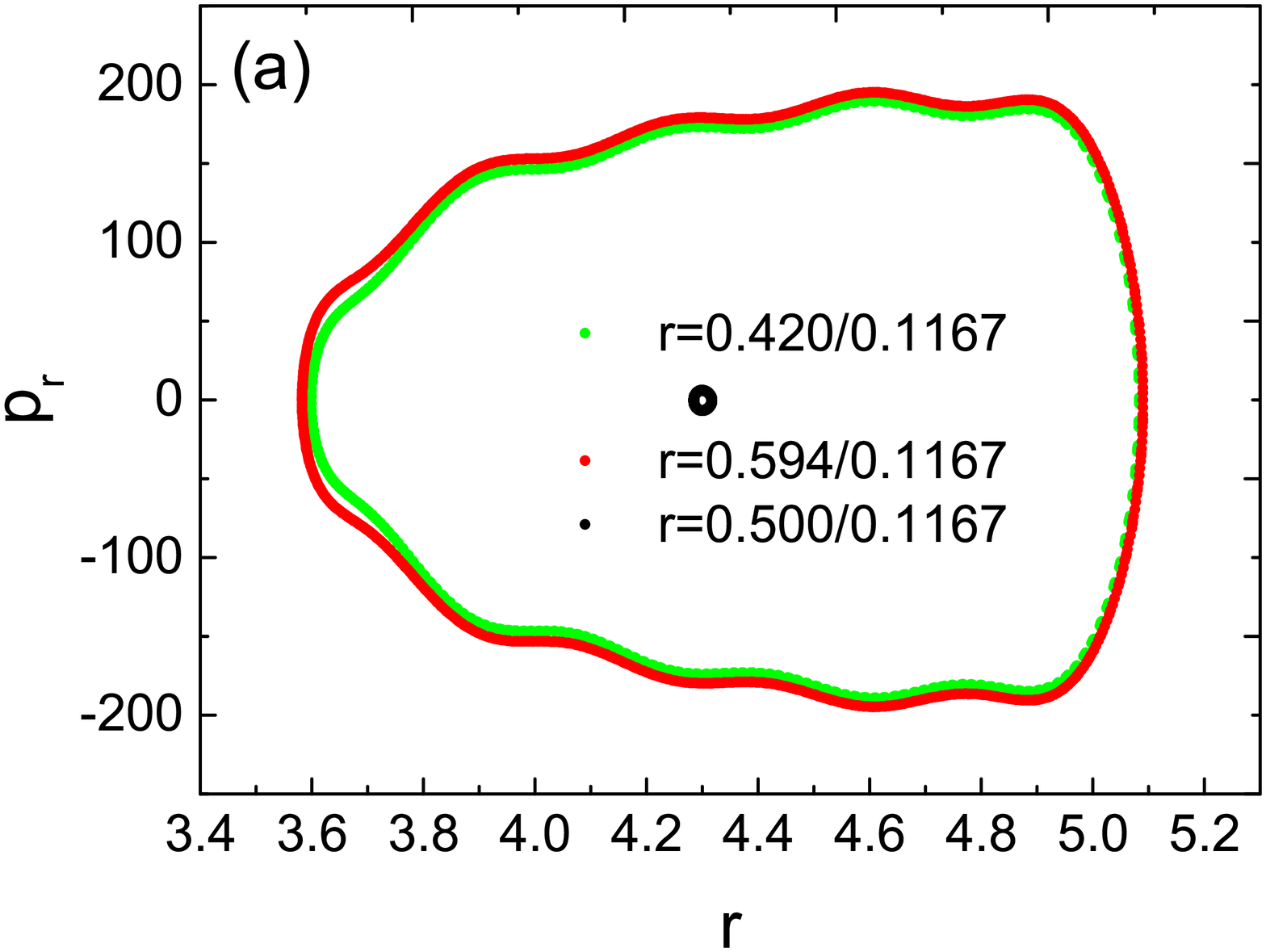}
\includegraphics[scale=0.20]{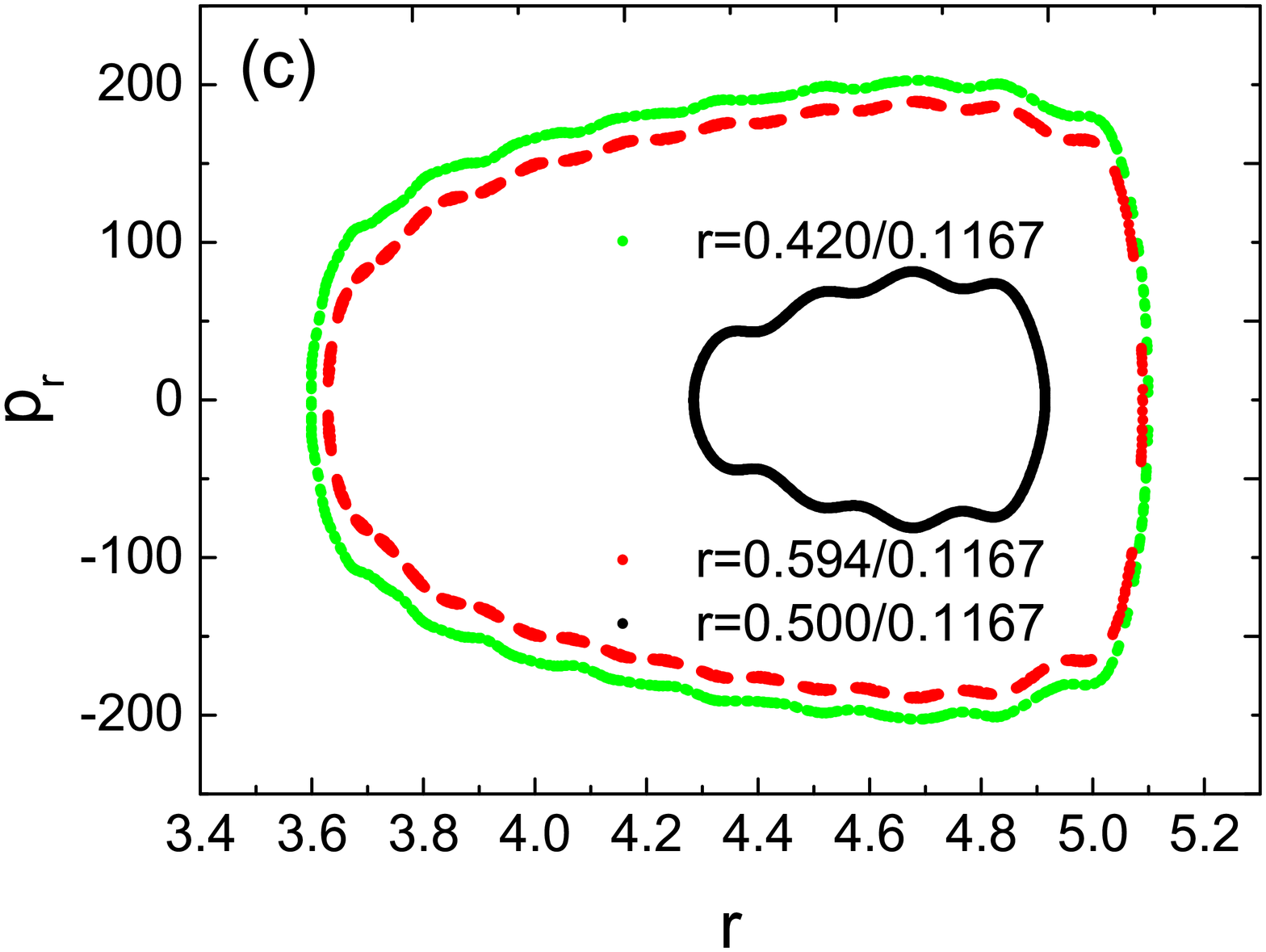}
\includegraphics[scale=0.19]{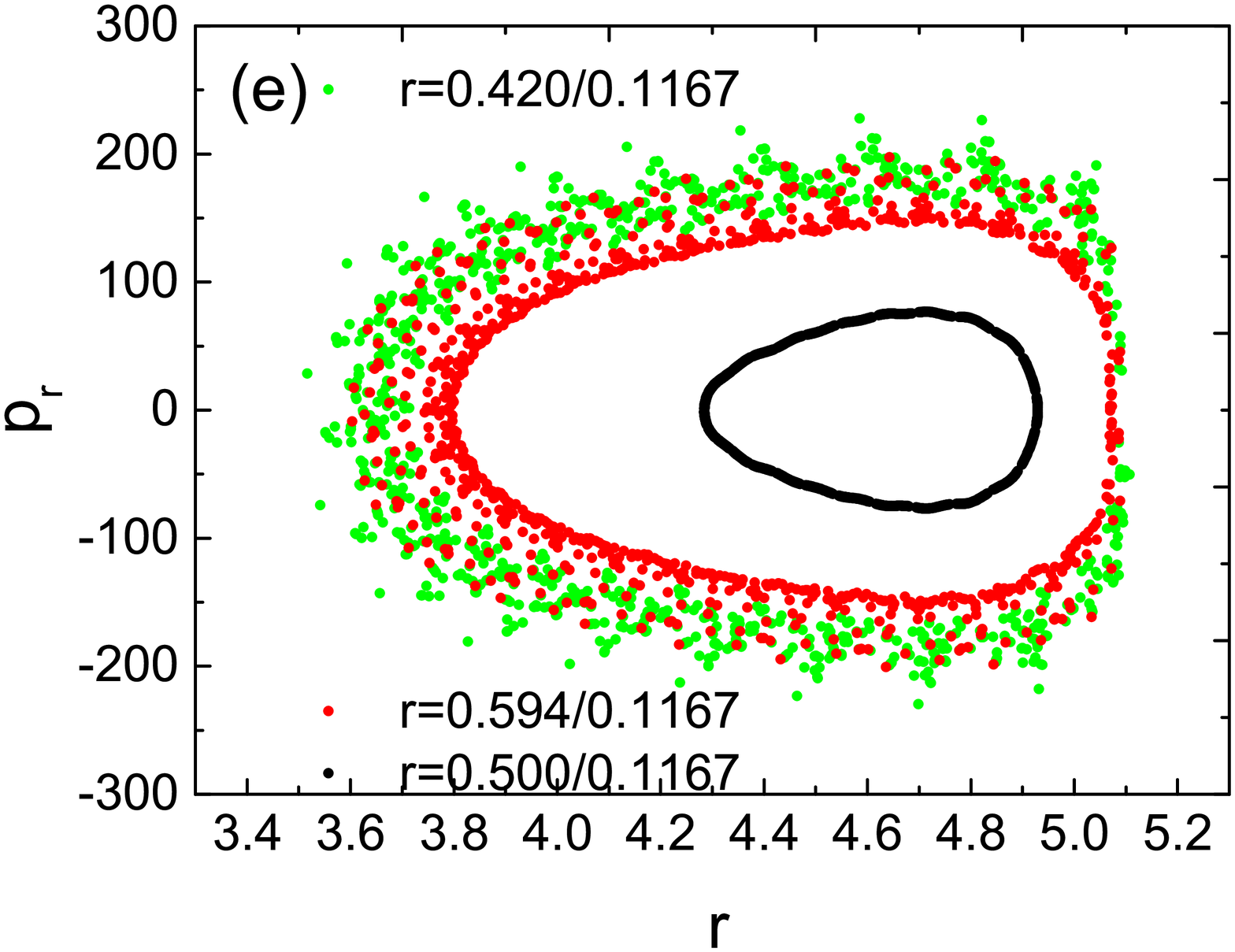}
\includegraphics[scale=0.20]{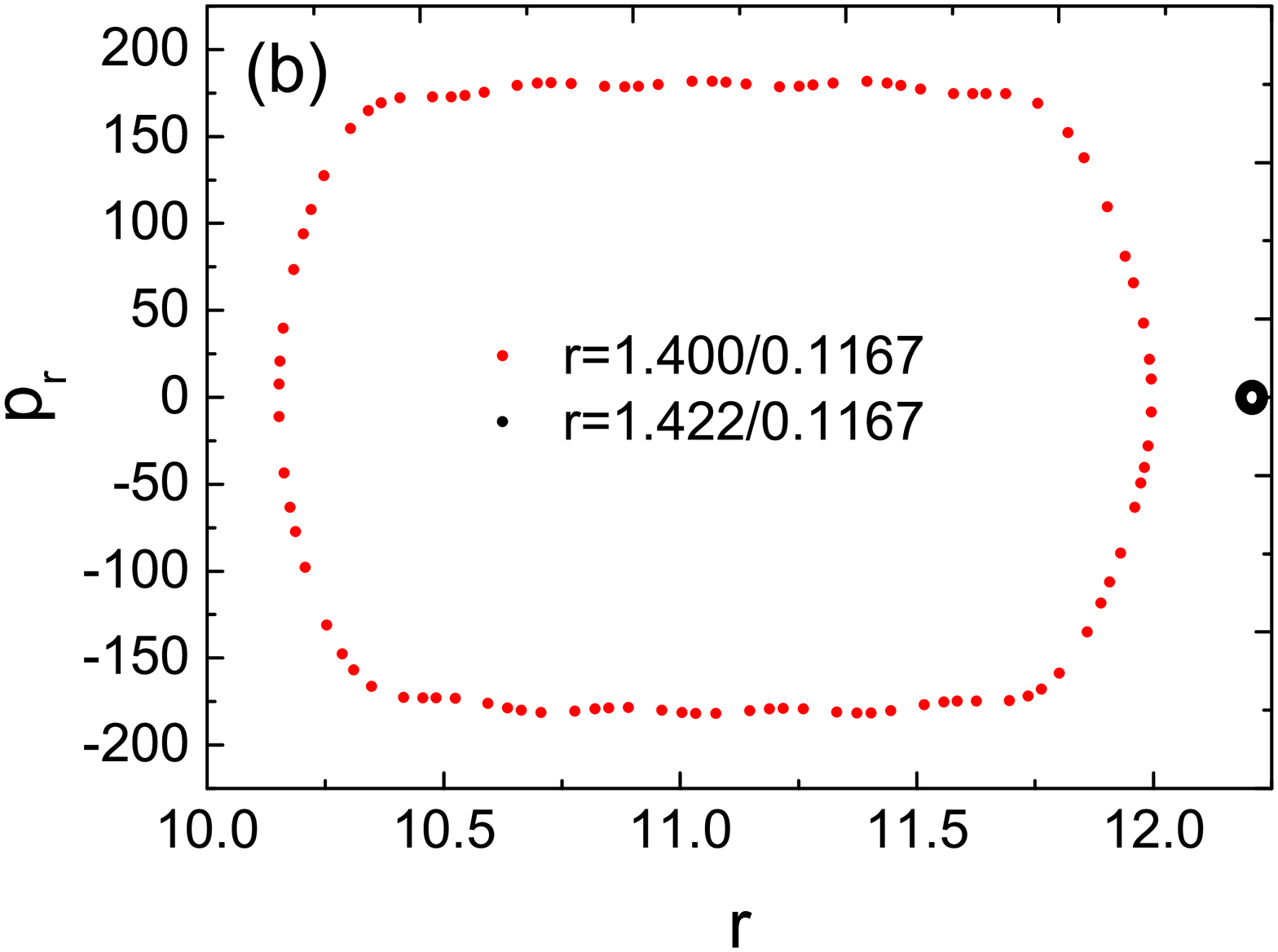}
\includegraphics[scale=0.20]{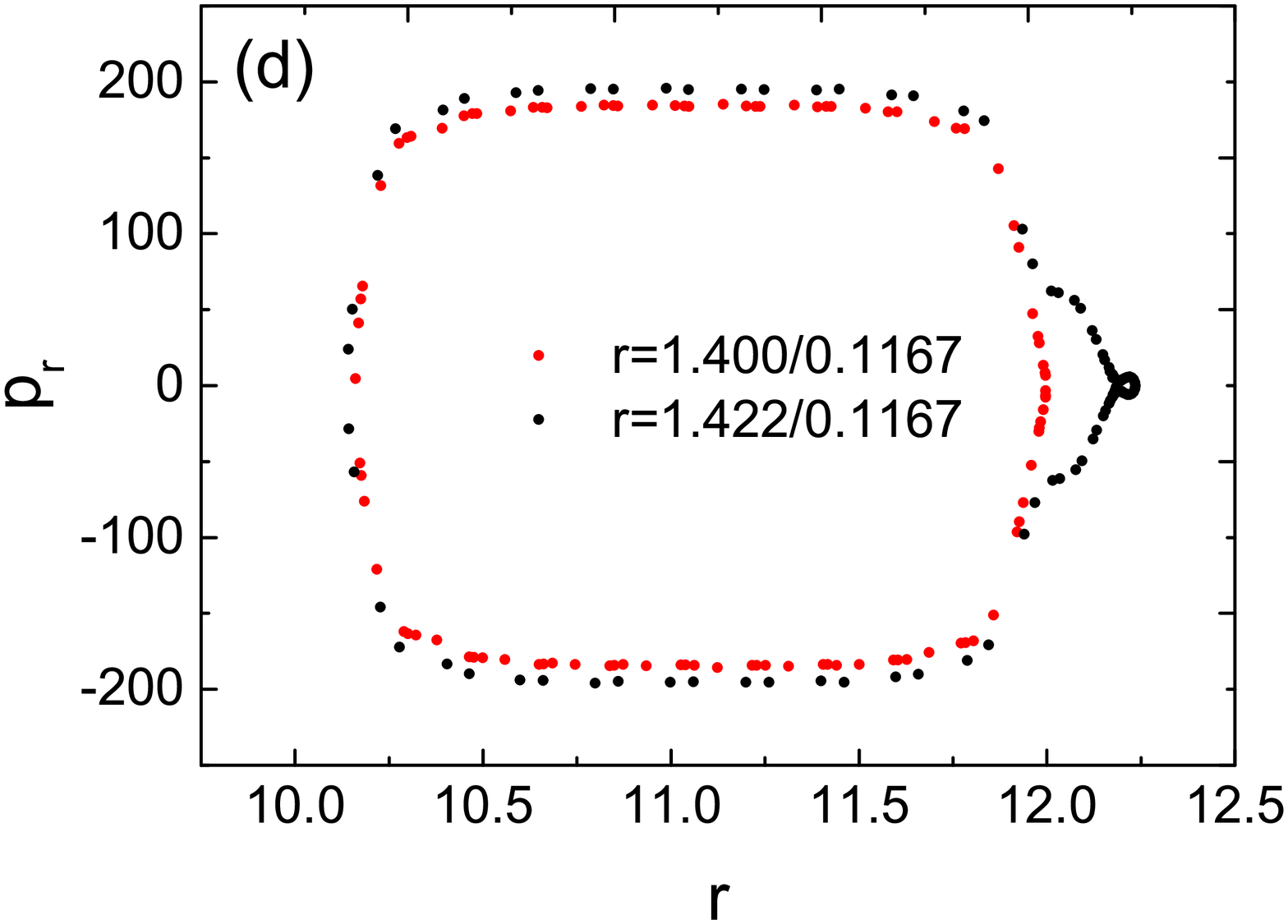}
\includegraphics[scale=0.20]{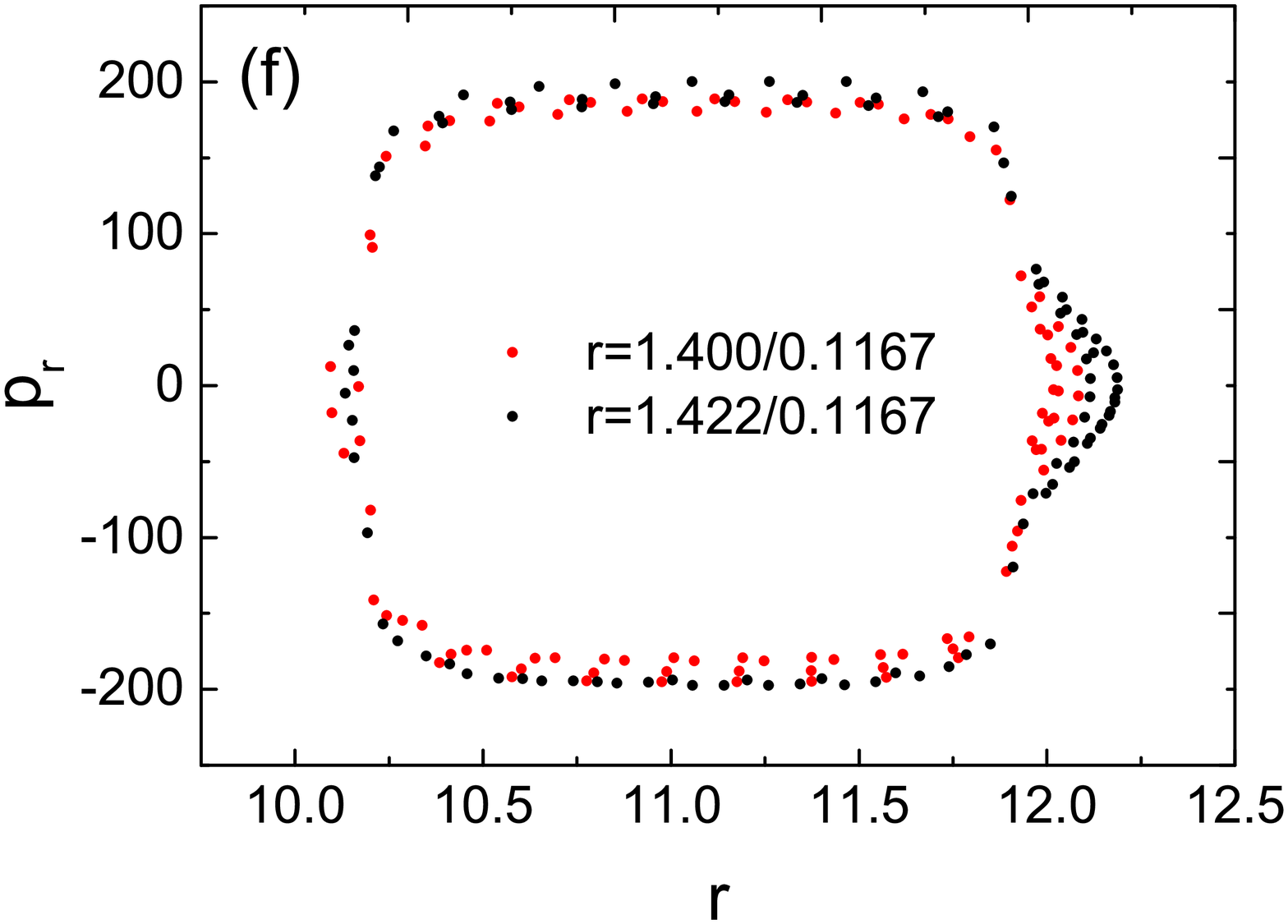}
\caption{(a) and (b): Phase space structures of Model C4 on the
Poincar\'{e} section. The structures are considered in the ranges
of $3.5<r<5.1$ and $11.9<r<12.2$. (c) and (d): Phase space
structures of Model C9 on the Poincar\'{e} section. (e) and (f):
Phase space structures of Model C19 on the Poincar\'{e} section.}}
\label{f10}
\end{figure*}

\begin{figure*}
\center{
\includegraphics[scale=0.20]{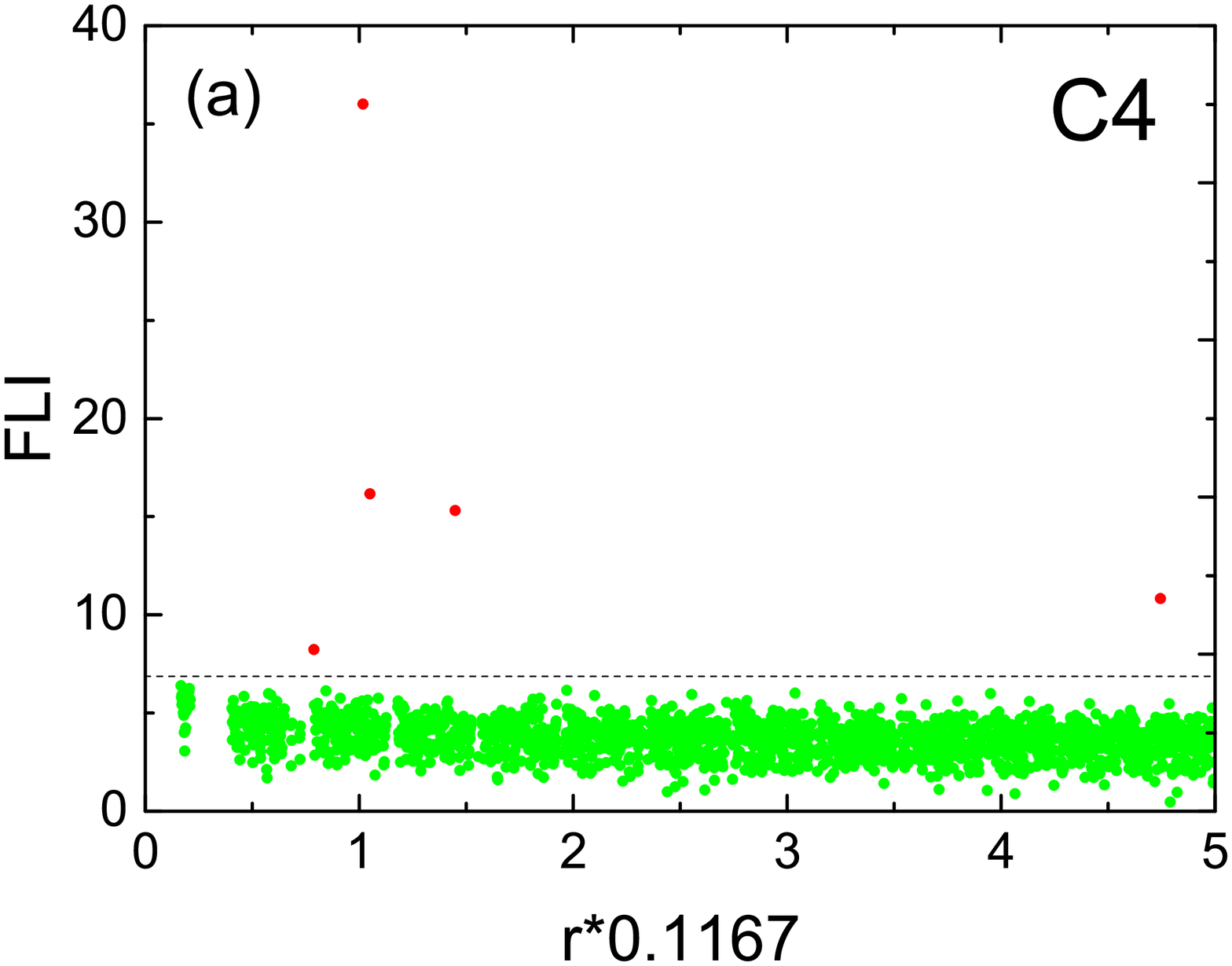}
\includegraphics[scale=0.20]{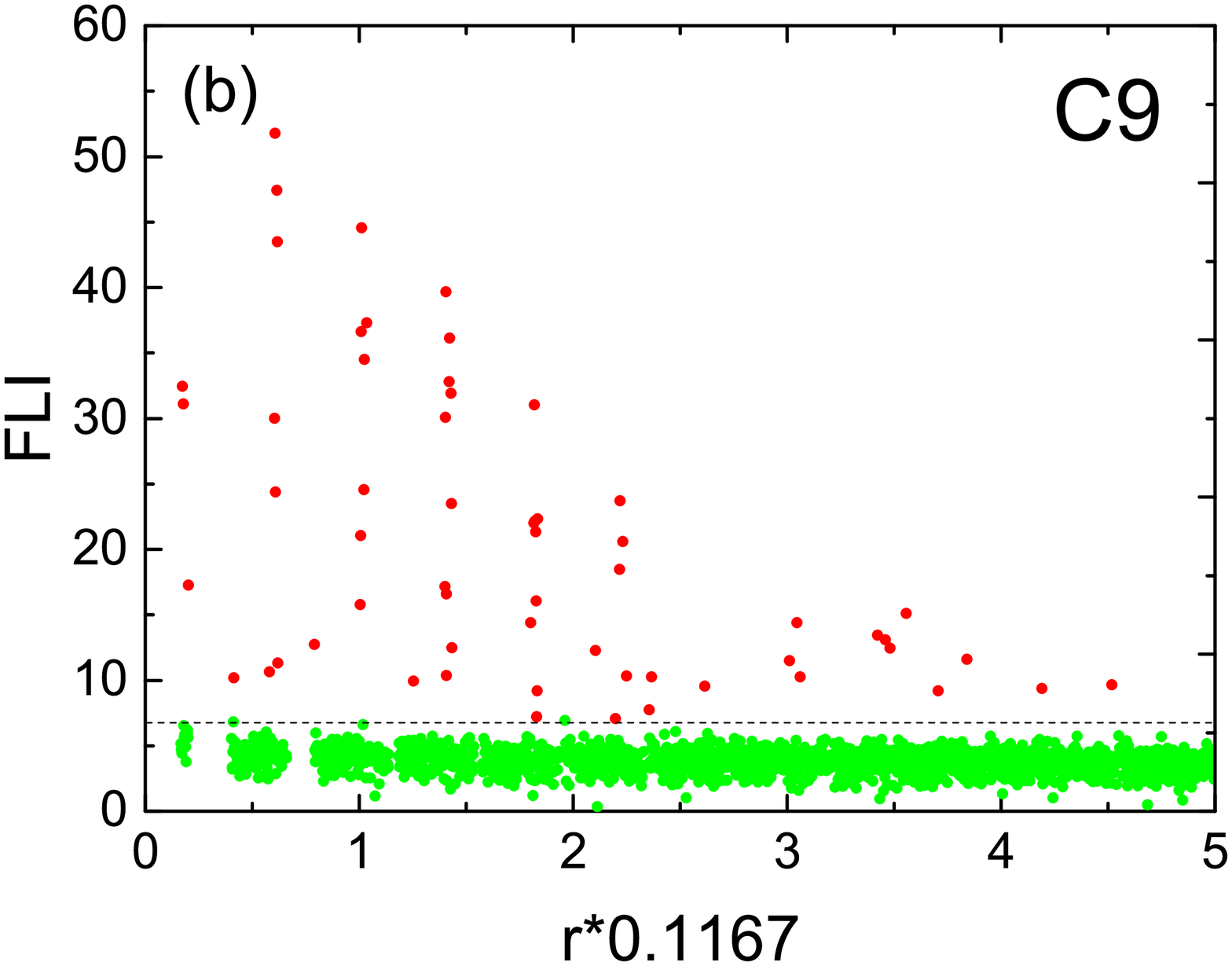}
\includegraphics[scale=0.20]{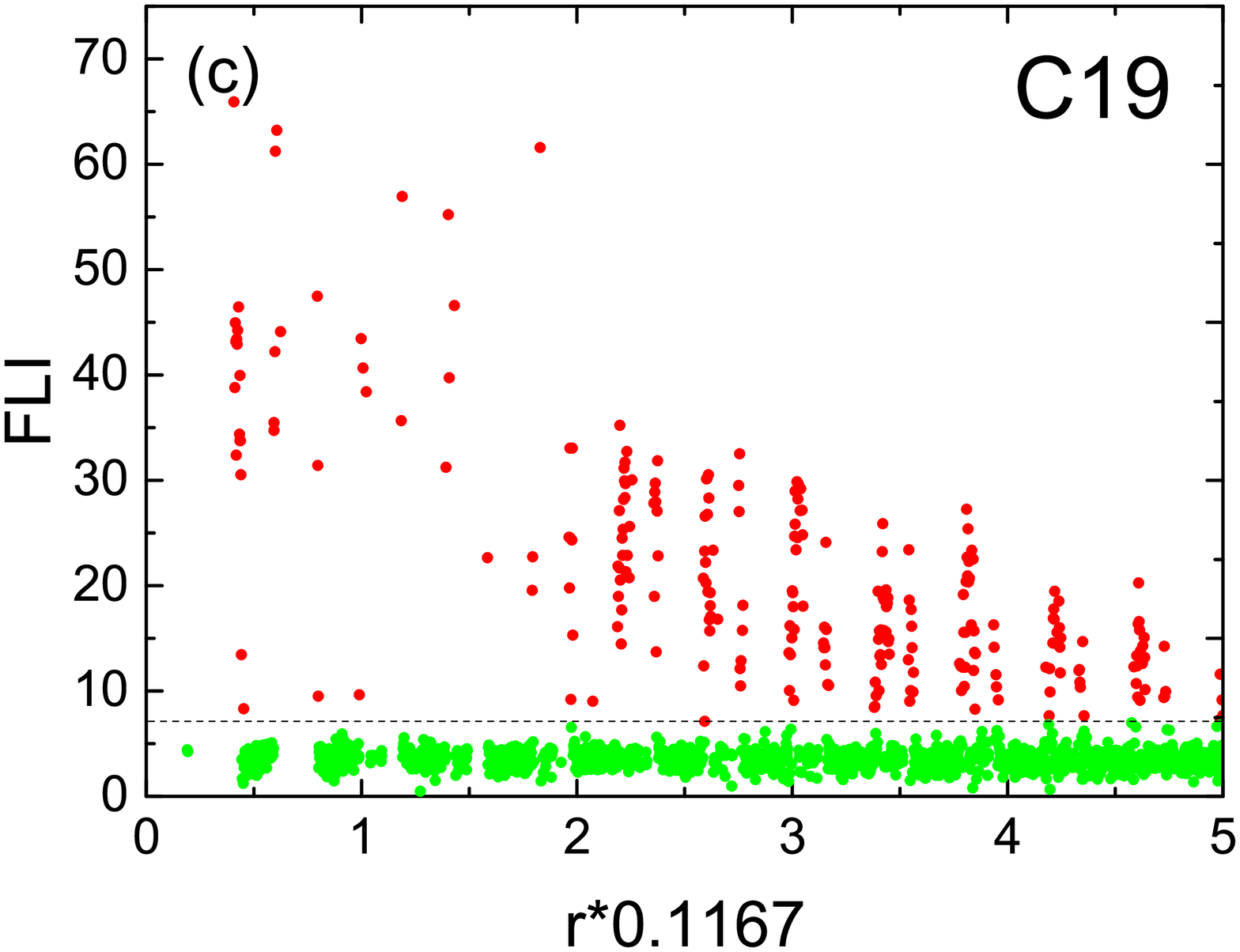}
\caption{Similar to Fig. 5 but Models A are replaced with Models
C. The degree of chaos increases with the radial term number $n$
increasing.}} \label{f11}
\end{figure*}

\end{document}